\documentclass{amsart}
\usepackage{amsmath}
\usepackage{amssymb}

\newtheorem{theorem}{Theorem}

\newtheorem{definition}[theorem]{Definition}

\newtheorem{lemma}[theorem]{Lemma}

\newtheorem{remark}[theorem]{Remark}

\begin{document}
\title[De Rham-Hodge-Skrypnik theory ]{A survey of the spectral and
differential geometric aspects of the De Rham-Hodge-Skrypnik theory related
with Delsarte transmutation operators in multidimension and its applications
to spectral and soliton problems. Part 2}
\thanks{The third author was supported in part by a local AGH grant.}
\author{Y.A. Prykarpatsky*)**)}
\address{*)The AGH University of Science and Technology, Department of
Applied Mathematics, Krakow 30059 Poland, and Brookhaven Nat. Lab., CDIC,
Upton, NY, 11973 USA}
\email{yarchyk@imath.kiev.ua, yarpry@bnl.gov}
\author{A.M. Samoilenko**)}
\address{**)The Institute of Mathematics, NAS, Kyiv 01601, Ukraine}
\author{A. K. Prykarpatsky***)}
\address{***)The AGH University of Science and Technology, Department of
Applied Mathematics, Krakow 30059 Poland}
\email{pryk.anat@ua.fm, prykanat@cybergal.com}

\begin{abstract}
The differential-geometric and topological structure of Delsarte
transmutation operators and associated with them Gelfand-Levitan-Marchenko
type eqautions are studied making use of the De Rham-Hodge-Skrypnik
differential complex. The relationships with spectral theory and special
Berezansky type congruence properties of Delsarte transmuted operators are
stated. Some applications to multidimensional differential operators are
done including three-dimensional Laplace operator, two-dimensional classical
Dirac operator and its multidimensional affine extension, related with
self-dual Yang-Mills eqautions. The soliton like solutions to the related
set of nonlinear dynamical systemare discussed.
\end{abstract}

\keywords{Delsarte transmutation operators, De Rham-Hodge-Skrypnik
differential complex, Darboux transformations, Gelfand-Levitan-Marchenko
equations, Dirac operator, Laplace operator, soliton like\ \ solutions}
\subjclass{Primary 34A30, 34B05 Secondary 34B15\\
PACS: 02.30.Jr, 02.30.Uu, 02.30.Zz, 02.40.Sf}
\maketitle



\setcounter{equation}{0}

\section{De Rham-Hodge-Skrypnik theory aspects and related \newline
Delsarte-Darboux type binary transformations}

\textbf{1.1} A differential-geometric analysis of Delsarte-Darboux type
transformations that was done in Chapter 3 of Part 1 for differential
operator expressions acting in a functional space $\mathcal{H}=L_{1}(\mathrm{%
T;}H)$, where $\mathrm{T}=\mathbb{R}^{2}$ and $H:=L_{2}(\mathbb{R}^{2};%
\mathbb{C}^{2})$, appears to have a deep relationship with classical de
Rham-Hodge-Skrypnik theory \cite{Sk,Sk1,Sk2,Sk3} devised in the midst of the
past century for a set of commuting differential operators defined, in
general on a smooth compact $m$-dimensional metric space $M.$ Concerning our
problem of describing the spectral structure of Delsarte-Darboux type
transmutaions acting in $\mathcal{H},$ we preliminarily consider following
Part 1 some backgrounds of the de-Rham-Hodge-Skrypnik differential complex
theory devised for studying transformations of differential operators.
Consider a smooth metric space $M$ being a suitably compactified form of the
space $\mathbb{R}^{m},$ $m\in \mathbb{Z}_{+}.$ Then one can define on $M_{%
\mathrm{T}}:=\mathrm{T}\times M$ \ the standard Grassmann algebra $\mathrm{%
\Lambda }(M_{\mathrm{T}};\mathcal{H})$ of differential forms on $\mathrm{%
T\times }M$ and consider a generalized external I.V. Skrypnik \cite{Sk,Sk1}
anti-differentiation operator $d_{\mathcal{L}}:\mathrm{\Lambda (}M\mathrm{_{%
\mathrm{T}}};\mathcal{H})\rightarrow \mathrm{\Lambda (}M_{\mathrm{T}};%
\mathcal{H})$ acting as follows: for any $\beta ^{(k)}\in \mathrm{\Lambda }%
^{k}(M_{\mathrm{T}};\mathcal{H}),$ $k=\overline{0,m},$
\begin{equation}
d_{\mathcal{L}}\beta ^{(k)}:=\sum_{j=1}^{2}dt_{j}\wedge \mathrm{L}%
_{j}(t;x|\partial )\beta ^{(k)}\in \mathrm{\Lambda }^{k+1}(M_{\mathrm{T}};%
\mathcal{H}),  \label{1.1}
\end{equation}%
where, by definition,
\begin{equation}
\mathrm{L}_{j}(t;x|\partial ):=\partial /\partial t_{j}-L_{j}(t;x|\partial )
\label{1.2}
\end{equation}%
$j=\overline{1,2},$ are suitably defined linear differential operators in $%
\mathcal{H}$, commuting to each other, that is
\begin{equation}
\lbrack \mathrm{L}_{1},\mathrm{L}_{2}]=0.  \label{1.3}
\end{equation}%
We will put, in general, that differential expressions
\begin{equation}
L_{j}(t;x|\partial ):=\sum_{|\alpha |=0}^{n_{j}(L)}a_{\alpha }^{(j)}(t;x)%
\frac{\partial ^{_{|\alpha |}}}{\partial x^{\alpha }},  \label{1.4}
\end{equation}%
with coefficients $a_{\alpha }^{(j)}\in C^{1}(\mathrm{T};C^{\infty }(M;End%
\mathbb{C}^{N})),$ $|\alpha |=\overline{0,n_{j}(L)}$ $n_{j}^{\alpha }\in
\mathbb{Z}_{+},$ $j=\overline{0,1},$ are some closed normal densely defined
operators in the Hilbert space $H$ \ for any $t\in \mathrm{T}$. It is easy
to observe that the anti-differentiation of $d_{\mathcal{L}}$ defined by (%
\ref{1.1}) is a generalization of the usual external anti-differentiation
\begin{equation}
d=\sum_{j=1}^{m}dx_{j}\wedge \frac{\partial }{\partial x_{j}}%
+\sum_{s=1}^{2}dt_{s}\wedge \frac{\partial }{\partial t_{s}}  \label{1.5}
\end{equation}%
for which, evidently, commutation conditions
\begin{equation}
\lbrack \frac{\partial }{\partial x_{j}};\frac{\partial }{\partial x_{k}}]=0,%
\text{ \ }[\frac{\partial }{\partial t_{s}};\frac{\partial }{\partial t_{l}}%
]=0,\text{ }[\frac{\partial }{\partial x_{j}};\frac{\partial }{\partial t_{s}%
}]=0  \label{1.6}
\end{equation}%
hold for all $j,k=\overline{1,m}$ and $s,l=\overline{1,2}.$ If now to
substitute within (\ref{1.5}) $\partial /\partial x_{j}\longrightarrow
\mathrm{A}_{j},$ $\partial /\partial t_{s}\longrightarrow \mathrm{L}_{s},$ $%
j=\overline{1,m},$ $s=\overline{1,2},$ one gets the anti-differentiation
\begin{equation}
d_{\mathcal{A}}:=\sum_{j=1}^{m}dx_{j}\wedge \mathrm{A}_{j}(t;x|\partial
)+\sum_{j=1}^{2}dt_{s}\wedge \mathrm{L}_{s}(t;x|\partial ),  \label{eq:1.7}
\end{equation}%
where the differential expressions $\mathrm{A}_{j},\mathrm{L}_{S}:\mathcal{H}%
\longrightarrow \mathcal{H}$ for all $j,k=\overline{1,m}$ and $s,l=\overline{%
1,2},$ satisfy the commutation conditions $[\mathrm{A}_{j},\mathrm{A}_{k}]=0$%
, $[\mathrm{L}_{s},\mathrm{L}_{s}]=0$, $[\mathrm{A}_{j},\mathrm{L}_{s}]=0,$
then then operation (\ref{eq:1.7}) defines on $\mathrm{\Lambda (}M\mathrm{_{%
\mathrm{T}}};\mathcal{H})$ an anti-differential with respect to which the
co-chain complex.
\begin{equation}
\mathcal{H}\longrightarrow \mathrm{\Lambda ^{0}(}M\mathrm{_{\mathrm{T}}};%
\mathcal{H})\overset{d_{\mathcal{A}}}{\longrightarrow }\mathrm{\Lambda ^{1}(}%
M\mathrm{_{\mathrm{T}}};\mathcal{H})\overset{d_{\mathcal{A}}}{%
\longrightarrow }...\overset{d_{\mathcal{A}}}{\longrightarrow }\mathrm{%
\Lambda ^{m+2}(}M\mathrm{_{\mathrm{T}}};\mathcal{H})\overset{d_{\mathcal{A}}}%
{\longrightarrow }0  \label{eq:1.8}
\end{equation}%
is evidently closed, that is $d_{\mathcal{A}}d_{\mathcal{A}}\equiv 0$. As
the anti-differential (\ref{1.1}) is a particular case of (\ref{eq:1.7}), we
obtain that the corresponding to it co-chain complex (\ref{eq:1.8}) is
closed too. \newline
\textbf{1.2} \ Below we will follow ideas developed \cite{Sk,Sk1,Sk2,Sk3,DeR}%
. A differential form $\beta \in \mathrm{\Lambda (}M\mathrm{_{\mathrm{T}}};%
\mathcal{H})$ will be called $d_{\mathcal{A}}$-closed if $d_{\mathcal{A}%
}\beta =0$ and a form $\gamma \in \mathrm{\Lambda (}M\mathrm{_{\mathrm{T}}};%
\mathcal{H})$ will be called exact or $d_{\mathcal{A}}$-homological to zero
if there exists on $M_{\mathrm{T}}$ such a form $\omega \in \mathrm{\Lambda (%
}M\mathrm{_{\mathrm{T}}};\mathcal{H})$ that $\gamma =d_{\mathcal{A}}\omega $.

Consider now the standard \cite{Te,DeR,Ch,Wa} algebraic Hodge star-operation
\begin{equation}
\ast :\mathrm{\Lambda ^{k}(}M\mathrm{_{\mathrm{T}}};\mathcal{H}%
)\longrightarrow \mathrm{\Lambda ^{m+2-k}(}M\mathrm{_{\mathrm{T}}};\mathcal{H%
}),  \label{eq:1.9}
\end{equation}%
$k=\overline{0,m+2}$, as follows: if $\beta \in \mathrm{\Lambda ^{k}(}M%
\mathrm{_{\mathrm{T}}};\mathcal{H}),$ then the form $\ast \beta \in \mathrm{%
\Lambda ^{m+2-k}(}M\mathrm{_{\mathrm{T}}};\mathcal{H})$ is such that:

\begin{itemize}
\item $(m-k+2)$ - dimensional volume $|\ast \beta |$ of the form $\ast \beta
$ equals $k$-dimensional volume $|\beta |$ of the form $\beta $;

\item the $(m+2)$ -dimensional measure $\bar{\beta}^{\mathrm{\intercal }%
}\wedge \ast \beta >0$ under the fixed orientation on $M_{\mathrm{T}}$.
\end{itemize}

Define also on the space $\mathrm{\Lambda (}M\mathrm{_{\mathrm{T}}};\mathcal{%
H})$ the following natural scalar product: for any $\beta ,\gamma \in
\mathrm{\Lambda ^{k}(}M\mathrm{_{\mathrm{T}}};\mathcal{H})$, $k=\overline{%
0,m,}$
\begin{equation}
(\beta ,\gamma ):=\int_{M_{\mathrm{T}}}\bar{\beta}^{\mathrm{\intercal }}\ast
\gamma .  \label{eq:1.10}
\end{equation}%
Subject to the scalar product (\ref{eq:1.10}) one can naturally construct
the corresponding Hilbert space
\begin{equation}
\mathcal{H}_{\Lambda }(M_{\mathrm{T}}):=\overset{m+2}{\underset{k=0}{\oplus }%
}\mathcal{H}_{\mathrm{\Lambda }}^{k}(M_{\mathrm{T}})  \label{eq:1.11}
\end{equation}%
well suitable for our further consideration. Notice also here, that the
Hodge star $\ast $-operation satisfies the following easily checkable
property: for any $\beta ,\gamma \in \mathcal{H}_{\mathrm{\Lambda }}^{k}(M_{%
\mathrm{T}})$, $k=\overline{0,m},$
\begin{equation}
(\beta ,\gamma )=(\ast \beta ,\ast \gamma ),  \label{eq:1.12}
\end{equation}%
that is the Hodge operation $\ast :\mathcal{H}_{\Lambda }(M_{\mathrm{T}%
})\rightarrow \mathcal{H}_{\Lambda }(M_{\mathrm{T}})$ is unitary and its
standard adjoint with respect to the scalar product (\ref{eq:1.10})
operation satisfies the condition $(\ast )^{\prime }=(\ast )^{-1}.$

Denote by $d_{\mathcal{L}}^{^{\prime }}$ the formally adjoint expression to
the weak differential operation (\ref{1.1}). By means of the operations $d_{%
\mathcal{L}}^{\prime }$ and $d_{\mathcal{L}}$ in the $\mathcal{H}_{\Lambda
}(M_{\mathrm{T}})$ one can naturally define \cite{Ch,Te,DeR,Sk,Wa} the
generalized Laplace-Hodge-Skrypnik operator $\Delta _{\mathcal{L}}:\mathcal{H%
}_{1}(M_{\mathrm{T}})\longrightarrow \mathcal{H}_{1}(M_{\mathrm{T}})$ as
\begin{equation}
\Delta _{\mathcal{L}}=d_{\mathcal{L}}^{\prime }d_{\mathcal{L}}+d_{\mathcal{L}%
}^{\prime }d_{\mathcal{L}}\text{ .}  \label{eq:1.13}
\end{equation}%
Take a form $\beta \in \mathcal{H}_{\Lambda }(M_{\mathrm{T}})$ satisfying
the equality
\begin{equation}
\Delta _{\mathcal{L}}\beta =0.  \label{eq:1.14}
\end{equation}%
Such a form is called \cite{Sk,DeR,Wa,Ch} harmonic. One can also verify that
a harmonic form $\beta \in \mathcal{H}_{\Lambda }(M_{\mathrm{T}})$ satisfies
simultaneously the following two adjoint conditions:
\begin{equation}
d_{\mathcal{L}}^{\prime }\beta =0,\quad d_{\mathcal{L}}\beta =0
\label{eq:1.15}
\end{equation}%
easily stemming from (\ref{eq:1.13}) and (\ref{eq:1.14}).

It is easy to check that the following differential operators in $\mathcal{H}%
_{\Lambda }(M_{\mathrm{T}})$
\begin{equation}
d_{\mathcal{L}}^{\ast }:=\ast d_{\mathcal{L}}^{\prime }(\ast )^{-1}
\label{eq:1.16}
\end{equation}%
defines also a new external anti-differential operation in $\mathcal{H}%
_{\Lambda }(M_{\mathrm{T}})$.

\begin{lemma}
The corresponding dual to (\ref{eq:1.8}) co-chain complex
\begin{equation}
\mathcal{H}\longrightarrow \mathrm{\Lambda }^{0}(M_{\mathrm{T}};\mathcal{H})%
\overset{d_{\mathcal{L}}^{\ast }}{\longrightarrow }\mathrm{\Lambda }^{1}(M_{%
\mathrm{T}};\mathcal{H})\overset{d_{\mathcal{L}}^{\ast }}{\longrightarrow }%
...\overset{d_{\mathcal{L}}^{\ast }}{\longrightarrow }\mathrm{\Lambda }%
^{m+2}(M_{\mathrm{T}};\mathcal{H})\overset{d_{\mathcal{L}}^{\ast }}{%
\longrightarrow }0  \label{eq:1.17}
\end{equation}%
is exact.
\end{lemma}

\begin{proof}
A proof follows owing to the property $d_{\mathcal{L}}^{\ast }d_{\mathcal{L}%
}^{\ast }=0$ holding due to the definition (\ref{eq:1.16}).$\triangleright $%
\newline
\end{proof}

\textbf{1.3} \ Denote further by $\mathcal{H}_{\Lambda (\mathcal{L})}^{k}(M_{%
\mathrm{T}}),$ $k=\overline{0,m+2},$ the cohomology groups of $d_{\mathcal{L}%
}$-closed and by $\mathcal{H}_{\mathrm{\Lambda }(\mathcal{L}^{\ast
})}^{k}(M_{\mathrm{T}})$, $k=\overline{0,m+2}$, $\ k=\overline{0,m+2},$ the
cohomology groups of $d_{\mathcal{L}}^{\ast }$-closed differential forms,
respectively, and by $\mathcal{H}_{\Lambda (\mathcal{L}^{\ast }\mathcal{L}%
)}^{k}(M_{\mathrm{T}})$, $k=\overline{0,m+2}$, the abelian groups of
harmonic differential forms from the Hilbert sub-spaces $\mathcal{H}_{%
\mathrm{\Lambda }}^{k}(M_{\mathrm{T}})$, $k=\overline{0,m+2}$. Before
formulating next results, define the standard Hilbert-Schmidt rigged chain
\cite{Be,BS} of positive and negative Hilbert spaces of differential forms
\begin{equation}
\mathcal{H}_{\mathrm{\Lambda },+}^{k}(M_{\mathrm{T}})\subset \mathcal{H}_{%
\mathrm{\Lambda }}^{k}(M_{\mathrm{T}})\subset \mathcal{H}_{\mathrm{\Lambda }%
,-}^{k}(M_{\mathrm{T}}),  \label{eq:1.18}
\end{equation}%
the corresponding hereditary rigged chains of harmonic forms:
\begin{equation}
\mathcal{H}_{\mathrm{\Lambda }(\mathcal{L}^{\ast }\mathcal{L}),+}^{k}(M_{%
\mathrm{T}})\subset \mathcal{H}_{\mathrm{\Lambda }(\mathcal{L}^{\ast }%
\mathcal{L})}^{k}(M_{\mathrm{T}})\subset \mathcal{H}_{\mathrm{\Lambda }(%
\mathcal{L}^{\ast }\mathcal{L}),-}^{k}(M_{\mathrm{T}})  \label{eq:1.19}
\end{equation}%
and chains of cohomology groups:
\begin{eqnarray}
\mathcal{H}_{\mathrm{\Lambda }(\mathcal{L}),+}^{k}(M_{\mathrm{T}}) &\subset &%
\mathcal{H}_{\mathrm{\Lambda }(\mathcal{L})}^{k}(M_{\mathrm{T}})\subset
\mathcal{H}_{\mathrm{\Lambda }(\mathcal{L}),-}^{k}(M_{\mathrm{T}}),
\label{eq:1.20} \\
\mathcal{H}_{\mathrm{\Lambda }(\mathcal{L}^{\ast }),+}^{k}(M_{\mathrm{T}})
&\subset &\mathcal{H}_{\mathrm{\Lambda }(\mathcal{L}^{\ast })}^{k}(M_{%
\mathrm{T}})\subset \mathcal{H}_{\mathrm{\Lambda }(\mathcal{L}^{\ast
}),-}^{k}(M_{\mathrm{T}})  \notag
\end{eqnarray}%
for all $k=\overline{0,m+2}$. Assume also that the Laplace-Hodge-Skrypnik
operator (\ref{eq:1.13}) is reduced upon the space $\mathcal{H}_{\mathrm{%
\Lambda }}^{0}(M)$. Now by reasoning similar to those in \cite{Ch,DeR,Wa}
one can formulate a little generalized \cite{Sk1,Sk2,Sk3,DeR} de
Rham-Hodge-Skrypnik theorem. \newline
The groups of harmonic forms $\mathcal{H}_{\Lambda ,+}^{k}(M_{\mathrm{T}}),$
$k=\overline{0,m+2},$ are, respectively, isomorphic to the homology groups $%
(H^{k}(M_{\mathrm{T}};\mathbb{C}))^{|\Sigma |},$ $k=\overline{0,m+2},$ where
$H^{k}(M_{\mathrm{T}};\mathbb{C})$ is the $k$-th cohomology group of the
manifold $M_{\mathrm{T}}$ with complex coefficients, a set $\Sigma \subset
\mathbb{C}^{p},$ $p\in \mathbb{Z}_{+},$ is the set of suitable "spectral"
parameters marking the linear space of independent $d_{\mathcal{L}}^{\ast }$%
-closed $0$-form from $\mathcal{H}_{\Lambda (\mathcal{L}),-}^{0}(M_{\mathrm{T%
}})$ and, moreover, the following direct sum decompositions
\begin{equation}
\mathcal{H}_{\Lambda ,+}^{k}(M_{\mathrm{T}})=\mathcal{H}_{\Lambda (\mathcal{L%
}^{\ast }\mathcal{L}),+}^{k}(M_{\mathrm{T}})\oplus \Delta _{\mathcal{L}}%
\mathcal{H}_{\Lambda ,+}^{k}(M_{\mathrm{T}})  \label{eq:1.21}
\end{equation}%
\begin{equation*}
=\mathcal{H}_{\Lambda (\mathcal{L}^{\ast }\mathcal{L}),+}^{k}(M\mathrm{_{T}}%
)\oplus d_{\mathcal{L}}\mathcal{H}_{\Lambda ,+}^{k-1}\mathrm{(}M\mathrm{_{T})%
}\oplus d_{\mathcal{L}}^{\prime }\mathcal{H}_{\Lambda ,+}^{k+1}(M_{\mathrm{T}%
})
\end{equation*}%
hold for any $k=\overline{0,m+2}.$

Another variant of the statement similar to that above was formulated in
\cite{Sk,Sk1} and reads as the following generalized de Rham-Hodge-Skrypnik
theorem. \newline

The generalized cohomology groups $\mathcal{H}_{\Lambda (\mathcal{L}%
),+}^{k}(M_{\mathrm{T}})$, $k=\overline{0,m+2},$ are isomorphic,
respectively, to the cohomology groups $(H^{k}(M_{\mathrm{T}};\mathbb{C}%
))^{|\Sigma |}$, $k=\overline{0,m+2}$.

A proof of this theorem is based on some special sequence \cite%
{Sk,Sk1,Sk2,Sk3,Lo} of differential Lagrange type identities.$\triangleright
$

Define the following closed subspace
\begin{equation}
\mathcal{H}_{0}^{\ast }:=\{\varphi ^{(0)}(\eta )\in \mathcal{H}_{\Lambda (%
\mathcal{L}^{\ast }),-}^{0}(M_{\mathrm{T}}):d_{\mathcal{L}}^{\ast }\varphi
^{(0)}(\eta )=0,\text{ }\varphi ^{(0)}(\eta )|_{\Gamma },\text{ }\eta \in
\Sigma \}  \label{eq:1.22}
\end{equation}%
for some smooth $(m+1)$-dimensional hypersurface $\Gamma \subset M_{\mathrm{T%
}}$ and $\Sigma \subset (\sigma (L)\cap \bar{\sigma}(L))\times \Sigma
_{\sigma }\subset \mathbb{C}^{p},$ where $\mathcal{H}_{\Lambda (\mathcal{L}%
^{\ast }),-}^{0}(M_{\mathrm{T}})$ is, as above, a suitable Hilbert-Schmidt
rigged\cite{Be,BS} zero-order cohomology group Hilbert space from the
co-chain given by (\ref{eq:1.20}), $\ \sigma (L)$ and $\sigma (L^{\ast })$
are, respectively, mutual generalized spectra of the sets of differential
operators $L$ and $L^{\ast }$ in $H$ at $t=0\in \mathrm{T}$. Thereby, the
dimension dim$\mathcal{H}_{0}^{\ast }=card$ $\Sigma :=|\Sigma |$ is assumed
to be known. The next lemma first stated by I.V. Skrypnik \cite{Sk,Sk1} is
of fundamental meaning for a proof of Theorem 1.2.

\begin{lemma}
There exists a set of differential $(k+1)$-forms $Z^{(k+1)}[\varphi
^{(0)}(\eta ),d_{\mathcal{L}}\psi ^{(k)}]$ $\in \Lambda ^{k+1}(M_{\mathrm{T}%
};\mathbb{C}),$ $k=$ $\overline{0,m+2},$ and a set of $k$-forms $%
Z^{(k)}[\varphi ^{(0)}(\eta ),\psi ^{(k)}]\in \Lambda ^{k}(M_{\mathrm{T}};%
\mathbb{C}),$ $k=\overline{0,m+2}$, parametrized by the set $\Sigma \ni \eta
,$being semilinear in $(\varphi ^{(0)}(\eta ),\psi ^{(k)})\in \mathcal{H}%
_{0}^{\ast }\times \mathcal{H}_{\Lambda ,+}^{k}(M_{\mathrm{T}})$, such that
\begin{equation}
Z^{(k+1)}[\varphi ^{(0)}(\eta ),d_{\mathcal{L}}\psi ^{(k)}]=dZ^{k}[\phi
^{(0)}(\eta \text{ }),\psi ^{(k)}]  \label{eq:1.23}
\end{equation}%
for all $k=\overline{0,m+2}$ and $\eta \in \Sigma .$
\end{lemma}

\begin{proof}
A proof is based on the following Lagrange type identity generalizing that
of Part 1 and holding for any pair $(\varphi ^{0}(\eta ),\psi ^{(}k))\in
\mathcal{H}_{0}^{\ast }\times \mathcal{H}_{\Lambda ,+}^{k}(M_{\mathrm{T}})$:
\begin{eqnarray}
0 &=&<d_{\mathcal{L}}^{\ast }\phi ^{_{(0)}}(\eta ),\ast (\psi ^{(k)}\wedge
\overline{\gamma })>=<\ast d_{\mathcal{L}}^{\prime }(\ast )^{-1}\varphi
^{(0)}(\eta ),\ast (\psi ^{(k)}\wedge \overline{\gamma })>  \label{eq:1.24}
\\
&=&<\ast d_{\mathcal{L}}^{\prime }(\ast )^{-1}\phi ^{(0)}(x),\psi
^{(k)}\wedge \overline{\gamma })>=  \notag \\
&=&<(\ast )^{-1}\varphi ^{(0)}(\eta ),d_{\mathcal{L}}\psi ^{(k)}\wedge
\overline{\gamma }>+Z^{(k+1)}[\psi ^{(0)}(\eta ),d_{\mathcal{L}}\psi {(k)}%
]\wedge \overline{\gamma }>=  \notag \\
&=&<(\ast )_{-1}\varphi ^{(0)}(\eta ),d_{\mathcal{L}}\psi ^{(k)}\wedge
\overline{\gamma }>+dZ^{(k)}[\varphi ^{(0)}(\eta ),\psi ^{(k)}]\wedge
\overline{\gamma },  \notag
\end{eqnarray}%
where $Z^{(k+1)}[\varphi ^{(0)}(\eta ),d_{\mathcal{L}}\psi ^{{(k)}}]\in
\Lambda ^{k+1}(M_{\mathrm{T}};\mathbb{C})$, $k=\overline{0,m+2}$, and $%
Z^{(k)}[\varphi ^{(0)}(\eta ),\psi ^{(k)}]\in \Lambda ^{k}(M_{\mathrm{T}};%
\mathbb{C})$, $k=\overline{0,m+2}$, are some semilinear differential forms
on $M_{\mathrm{T}}$ parametrized by a parameter $\lambda \in \Sigma ,$ and $%
\overline{\gamma }\in \Lambda ^{m+1-k}(M_{\mathrm{T}};\mathbb{C})$ is
arbitrary constant $(m+1-k)$-form. Thereby, the semilinear differential $%
(k+1)$-forms $Z^{(k+1)}[\varphi ^{(0)}(\eta ),d_{\mathcal{L}}\psi ^{{(k)}%
}]\in \Lambda ^{k+1}(M_{\mathrm{T}};\mathbb{C})$ and $k$-forms $%
Z^{(k)}[\varphi ^{(0)}(\eta ),\psi ^{(k)}]\in \Lambda ^{k}(M_{\mathrm{T}};%
\mathbb{C}),$ $\ k=\overline{0,m+2},$ $\lambda \in \Sigma ,$ constructed
above exactly constitute those searched for in the Lemma.$\triangleright $
\end{proof}

\bigskip

\textbf{1.4} Based now on Lemma 1.3 one can construct the cohomology group
isomorphism claimed in Theorem 1.2 formulated above. Namely, following \cite%
{Sk,Sk1}, let us take some singular simplicial \cite{Te,DeR,DeR1,Wa} complex
$\mathcal{K}(M_{\mathrm{T}})$ of the compact metric space $M_{\mathrm{T}}$
and introduce a set of linear mappings $B_{\lambda }^{(k)}:\mathcal{H}%
_{\Lambda ,+}^{k}M_{\mathrm{T}}$ $\longrightarrow C_{k}(M_{\mathrm{T}};%
\mathbb{C}),$ $k=\overline{0,m+2,}$ $\lambda \in \Sigma ,$ \ where $C_{k}(M_{%
\mathrm{T}};\mathbb{C}),$ $k=\overline{0,m+2,}$ are free abelian groups over
the field $\mathbb{C}$ generated, respectively, by all $k$-chains of
singular simplexes $S^{(k)}\subset M_{\mathrm{T}},$ $\ k=\overline{0,m+2}$,
from the simplicial complex $\mathcal{K}(M_{\mathrm{T}}),$ as follows:
\begin{equation}
B_{\lambda }^{(k)}(\psi ^{(k)}):=\sum_{S^{(k)}\in C_{k}(M_{\mathrm{T}};%
\mathbb{C}))}S^{(k)}\int_{S^{(k)}}Z^{(k)}[\varphi ^{(0)}(\lambda ),\psi
^{(k)}]  \label{eq:1.25}
\end{equation}%
with $\psi ^{{(k)}}\in \mathcal{H}_{\Lambda ,+}^{k}(M_{\mathrm{T}})$, $k=%
\overline{0,m+2}$. The following theorem \cite{Sk,Sk1} based on mappings (%
\ref{eq:1.25}) holds.\newline

\begin{theorem}
The set of operators (\ref{eq:1.25}) parametrized by $\lambda \in \Sigma $ \
realizes the cohomology group isomorphism formulated in Theorem 1.2
\end{theorem}

\begin{proof}
A proof of this theorem one can get passing over in (\ref{eq:1.25} ) to the
corresponding cohomology $\mathcal{H}_{\Lambda (\mathcal{L}),+}^{k}(M_{%
\mathrm{T}})$ and homology $H_{k}(M_{\mathrm{T}};\mathbb{C})$ groups of $M_{%
\mathrm{T}}$ for every $k=\overline{0,m+2}$. If one to take an element $\psi
^{(k)}:=\psi ^{(k)}(\mu )\in \mathcal{H}_{\Lambda (\mathcal{L}),+}^{k}(M_{%
\mathrm{T}})$, $k=\overline{0,m+2}$, solving the equation $d_{\mathcal{L}%
}\psi ^{(k)}(\mu )=0$ with $\mu \in \Sigma _{k}$ being some set of the
related "spectral" parameters marking elements of the subspace $\mathcal{H}%
_{\Lambda (\mathcal{L}),-}^{k}(M_{\mathrm{T}})$, then one finds easily from (%
\ref{eq:1.25}) and identity (\ref{eq:1.23}) that $dZ^{(k)}[\varphi
^{(0)}(\lambda ),$ $\psi ^{(k)}(\mu )]=0$ for all $(\lambda ,\mu )\in \Sigma
\times \Sigma _{k}$, $k=\overline{0,m+2}.$ This, in particular, means due to
the Poincare lemma \cite{Go,Te,DeR} \ that there exist differential $(k-1)$%
-forms $\Omega ^{(k-1)}[\varphi ^{(0)}(\lambda ),\psi ^{(k)}(\mu ]\in $ $%
\Lambda ^{k-1}(M;\mathbb{C})$, $k=\overline{0,m+2},$ such that
\begin{equation}
Z^{(k)}[\varphi ^{(0)}(\lambda ),\psi ^{(k)}(\mu )]=d\Omega ^{(k-1)}[\varphi
^{(0)}(\lambda ),\psi ^{(k)}(\mu )]  \label{eq:1.26}
\end{equation}%
for all pairs $(\varphi ^{(0)}(\lambda ),\psi ^{(k)}(\mu ))\in \mathcal{H}%
_{0}^{\ast }\times \mathcal{H}_{\Lambda (\mathcal{L}),+}^{k}(M_{\mathrm{T}})$
parametrized by $(\lambda ,\mu )\in \Sigma \times \Sigma _{k}$, $k=\overline{%
0,m+2}$. As a result of passing on the right hand-side of (\ref{eq:1.25}) to
the homology groups $H_{k}(M_{\mathrm{T}};\mathbb{C}),$ $k=\overline{0,m+2},$
one gets due to the standard Stokes theorem \cite{Go,DeR,Te} that the
mappings
\begin{equation}
B_{\lambda }^{(k)}:\mathcal{H}_{\Lambda (\mathcal{L}),+}^{k}(M_{\mathrm{T}%
})\longrightarrow H_{k}(M_{\mathrm{T}};\mathbb{C})  \label{eq:1.27}
\end{equation}%
are isomorphisms for every $k=\overline{0,m+2}$ and $\lambda \in \Sigma $.
Making further use of the Poincare duality \cite{Ch,Te,DeR} between the
homology groups $H_{k}(M_{\mathrm{T}};\mathbb{C}),$ $k=\overline{0,m+2}$,
and the cohomology groups $H^{k}(M;\mathbb{C})$, $k=\overline{0,m+2},$
respectively, one obtains finally the statement claimed in Theorem 1.4.$%
\triangleright $\newline
\end{proof}

\section{\protect\bigskip The spectral structure of Delsarte-Darboux type
transmutation operators in multidimension}

\textbf{2.1 } Take now into account that our differential operators $\mathrm{%
L}_{j}:\mathcal{H}\rightarrow \mathcal{H},$ $j=\overline{1,2},$ are of the
special form (\ref{1.2}). Assume also that differential expressions (\ref%
{1.4}) are normal closed operators defined on dense subspace $D(L)\subset
L_{2}(M;\mathbb{C}^{N})$.

Then due to Theorem 1.4 one can find such a pair $(\varphi ^{(0)}(\lambda
),\psi ^{(k)}(\mu ))\in \mathcal{H}_{0}^{\ast }\times \mathcal{H}_{\Lambda (%
\mathcal{L}),+}^{k}(M_{\mathrm{T}})$ parametrized by elements $(\lambda ,\mu
)\in \Sigma \times \Sigma _{k},$ for which the equality
\begin{equation}
B_{\lambda }^{(m)}(\psi ^{(0)}(\mu )dx)=S_{(t;x)}^{(m)}\int_{\partial
S_{(t;x)}^{(m)}}\Omega ^{(m-1)}[\varphi ^{(0)}(\lambda ),\psi ^{(0)}(\mu )dx]
\label{eq:2.1}
\end{equation}%
holds, where $S_{(t;x)}^{(m)}\in \mathrm{\ }H_{m}(M_{\mathrm{T}};\mathbb{C})$
is some arbitrary but fixed element of parametrized by an arbitrarily chosen
point $(t;x)\in M_{\mathrm{T}}\cap \partial S_{(t;x)}^{(m)}.$ Consider the
next integral expressions
\begin{eqnarray}
\Omega _{(t;x)}(\lambda ,\mu ) &:&=\int_{\sigma _{(t;x)}^{(m-1)}}\Omega
^{(m-1)}[\varphi ^{(0)}(\lambda ),\psi ^{(0)}(\mu )dx],  \label{eq:2.2} \\
\Omega _{(t_{0};x_{0})}(\lambda ,\mu ) &:&=\int_{\sigma
_{(t_{0};x_{0})}^{(m-1)}}\Omega ^{(m-1)}[\phi ^{(0)}(\lambda ),psi_{(0)}(\mu
)dx],  \notag
\end{eqnarray}%
with a point $(t_{0};x_{0})\in M_{\mathrm{T}}\cap \partial
S_{(t_{0};x_{0})}^{(m)}$ being taken fixed the boundaries $\sigma
_{(t;x)}^{(m-1)}:=\partial S_{t;x}^{(m)},$ $\sigma
_{(t_{0};x_{0})}^{(m-1)}:=\partial S_{t_{0};x_{0}}^{(m)}$ assumed to be
homological to each other as $(t;x_{0})\longrightarrow (t;x)\in M_{\mathrm{T}%
},$ $(\lambda ,\mu )\in \Sigma \times \Sigma _{k},$ and interpret them as
the kernels \cite{Be,BS,DS} of the corresponding invertible integral
operators of Hilbert-Schmidt type $\Omega _{(t;x)},\Omega
_{(t_{0};x_{0})}:L_{2}^{(\rho )}(\Sigma ;\mathbb{C})\longrightarrow
L_{2}^{(\rho )}(\Sigma ;\mathbb{C}),$ where $\rho $ is some finite Borel
measure on the parameters\ set $\Sigma .$ Define now the invertible
operators expressions
\begin{equation}
\mathbf{\Omega }_{\pm }:\psi ^{(0)}(\mu )\longrightarrow \tilde{\psi}%
^{(0)}(\mu )  \label{eq:2.3}
\end{equation}%
for $\psi ^{(0)}(\mu )dx\in \mathcal{H}_{\Lambda (\mathcal{L}),+}^{m}(M_{%
\mathrm{T}})$ and some $\tilde{\psi}^{(0)}(\mu )dx\in \mathcal{H}_{\Lambda (%
\mathcal{L}),+}^{m}(M_{\mathrm{T}}),$ $\mu \in \Sigma ,$ where, by
definition, for any $\eta \in \Sigma $
\begin{eqnarray}
\tilde{\psi}^{(0)}(\eta ) &:&=\psi ^{(0)}(\eta )\cdot \Omega
_{(t;x)}^{-1}\cdot \Omega _{(t_{0};x_{0})}  \label{eq:2.4} \\
&=&\int_{\Sigma }d\rho (\mu )\int_{\Sigma }d\rho (\xi )\psi ^{(0)}(\mu
)\Omega _{(t;x)}^{-1}(\mu ,\xi )\Omega _{(t_{0};x_{0})}(\xi ,\eta ),  \notag
\end{eqnarray}%
being motivated by the expression (\ref{eq:2.1}). Namely, consider the
following diagram \newline
\begin{equation}
\begin{array}{ccc}
\mathcal{H}_{\Lambda (\mathcal{L}),+}^{m}(M_{\mathrm{T}}) & \overset{\Omega
_{\pm }}{\longrightarrow } & \mathcal{H}_{\Lambda (\mathcal{\tilde{L}}%
),+}^{m}(M_{\mathrm{T}}), \\
B_{\lambda }^{(m)}\downarrow & \swarrow \tilde{B}_{\lambda }^{(m)} &  \\
H_{m}(M_{\mathrm{T}};\mathbb{C}) &  &
\begin{array}{c}
\\
\end{array}%
\end{array}
\label{eq:2.5}
\end{equation}%
\newline
which is assumed to be commutative for some another co-chain complex
\begin{equation}
\mathcal{H}\longrightarrow \mathrm{\Lambda }^{0}(M_{\mathrm{T}};\mathcal{H})%
\overset{d_{\mathcal{\tilde{L}}}}{\longrightarrow }\mathrm{\Lambda }^{1}(M_{%
\mathrm{T}};\mathcal{H})\overset{d_{\mathcal{\tilde{L}}}}{\longrightarrow }%
...\overset{d_{\mathcal{\tilde{L}}}}{\longrightarrow }\mathrm{\Lambda }%
^{m+2}(M_{\mathrm{T}};\mathcal{H})\overset{d_{\mathcal{\tilde{L}}}}{%
\longrightarrow }0.  \label{eq:2.6}
\end{equation}%
Here, by definition, the generalized anti-differentiation is
\begin{equation}
d_{\tilde{\mathcal{L}}}:=\sum_{j=1}^{2}dt_{j}\wedge \tilde{\mathrm{L}}%
_{j}(t;x|\partial )  \label{eq:2.7}
\end{equation}%
with
\begin{eqnarray}
\tilde{\mathrm{L}}_{j} &=&\partial /\partial t_{j}-\tilde{L}%
_{j}(t;x|\partial ),  \label{eq:2.8} \\
\tilde{L}_{j}(t;x|\partial ) &:&=\sum_{|\alpha |=0}^{n_{j}(\tilde{L})}\tilde{%
a}_{\alpha }^{(j)}(t;x)\frac{\partial ^{|\alpha |}}{\partial x^{\alpha }},
\notag
\end{eqnarray}%
where coefficients $\tilde{a}_{\alpha }^{(j)}\in C^{1}(\mathrm{T};C^{\infty
}(M;\mathrm{End}\mathbb{C}^{N})$, $|\alpha |=\overline{0,n_{j}(\tilde{L)}},$
$\ n_{j}(\tilde{L}):=n_{j}(L)\in \mathbb{Z}_{+},$ $j=\overline{1,2}.$ The
corresponding isomorphisms $\tilde{B}_{\lambda }^{(m)}:\mathcal{H}_{\Lambda (%
\mathcal{L}),+}^{m}(M_{\mathrm{T}})\longrightarrow H_{m}(M_{\mathrm{T}};%
\mathbb{C}),$ $\lambda \in \Sigma ,$ act, by definition, as follows:
\begin{equation}
\tilde{B}_{\lambda }^{(m)}(\tilde{\psi}^{(0)}(\mu
)dx)=S_{(t;x)}^{(m)}\int_{\partial S_{(t;x)}^{(m)}}\tilde{\Omega}^{(m-1)}[%
\tilde{\varphi}^{(0)}(\lambda ),\tilde{\psi}^{(0)}(\mu )dx],  \label{eq:2.9}
\end{equation}%
where $\tilde{\varphi}^{(0)}(\lambda )\in \mathcal{\tilde{H}}_{0}^{\ast
}\subset \mathcal{H}_{\Lambda (\mathcal{L}^{\ast }),-}^{0}(M_{\mathrm{T}}),$
$\ \lambda \in (\sigma (\tilde{L})\cap \bar{\sigma}(\tilde{L}^{\ast
}))\times \Sigma _{\sigma },$
\begin{equation}
\mathcal{\tilde{H}}_{0}^{\ast }:=\{\tilde{\varphi}^{(0)}(\lambda )\in
\mathcal{H}_{\Lambda (\mathcal{L}^{\ast }),-}^{m}(M_{\mathrm{T}}):d_{\tilde{%
\mathcal{L}}}^{\ast }\tilde{\varphi}^{(0)}(x)=0,\tilde{\varphi}%
^{(0)}(\lambda )|_{\tilde{\Gamma}}=0,\lambda \in \Sigma \}  \label{eq:2.10}
\end{equation}%
for some hypersurface $\tilde{\Gamma}\subset \mathrm{\ }M_{\mathrm{T}}$.
Respectively, one defines the following closed subspace
\begin{equation}
\mathcal{\tilde{H}}_{0}:=\{\tilde{\psi}^{(0)}(\mu )\in \mathcal{H}_{\Lambda (%
\mathcal{L}^{\ast }),-}^{0}(M_{\mathrm{T}}):d_{\tilde{\mathcal{L}}}^{\ast }%
\tilde{\psi}^{(0)}(\lambda )=0,\tilde{\psi}^{(0)}(\mu )|_{\tilde{\Gamma}%
}=0,\mu \in \Sigma \}  \label{eq:2.11}
\end{equation}%
for the hyperspace $\tilde{\Gamma}\subset \mathrm{\ }M_{\mathrm{T}},$
introduced above. \newline
Suppose now that the elements (\ref{eq:2.4}) belong to the closed subspace (%
\ref{eq:2.11}), that is
\begin{equation}
d_{\tilde{\mathcal{L}}}\tilde{\psi}^{(0)}(\mu )=0  \label{eq:2.12}
\end{equation}%
Define similarly to (\ref{eq:2.11}) a closed subspace $\mathcal{\tilde{H}}%
_{0}^{\ast }\subset \mathcal{H}_{\Lambda (\mathcal{L}^{\ast }),-}^{m}(M_{%
\mathrm{T}})$ as follows:
\begin{equation}
\mathcal{H}_{0}:=\{\psi ^{(0)}(\lambda )\in \mathcal{H}_{\Lambda (\mathcal{L}%
^{\ast }),-}^{0}(M_{\mathrm{T}}):d_{\mathcal{L}}\psi ^{(0)}(\lambda )=0,\psi
^{(0)}(\lambda )|_{\Gamma }=0,\lambda \in \Sigma \}  \label{eq:2.13}
\end{equation}%
for all $\mu \in \Sigma $. Then due to the commutativity of the diagram (\ref%
{eq:2.5}) there exist the corresponding two invertible mappings
\begin{equation}
\mathbf{\Omega }_{\pm }:\mathcal{H}_{0}\rightarrow \mathcal{\tilde{H}}_{0},
\label{eq:2.14}
\end{equation}%
depending on ways of their extending over the whole Hilbert space $\mathcal{H%
}_{\Lambda ,-}^{m}(M_{\mathrm{T}}).$ Extend now operators (\ref{eq:2.14})
upon the whole Hilbert space $\mathcal{H}_{\Lambda ,-}^{m}(M_{\mathrm{T}})$
by means of the standard method \cite{PSP,SP} of variation of constants,
taking into account that for kernels $\Omega _{(t;x)}(\lambda ,\mu ),\Omega
_{(t_{0};x_{0})}(\lambda ,\mu )$ $\in L_{2}^{(p)}(\Sigma ;\mathbb{C})\otimes
L_{2}^{(p)}(\Sigma ;\mathbb{C}),$ $\lambda ,\mu \in \Sigma ,$ one can write
down the following relationships:
\begin{eqnarray}
&&\Omega _{(t;x)}(\lambda ,\mu )-\Omega _{(t_{0};x_{0})}(\lambda ,\mu )
\label{eq:2.15} \\
&=&\int_{\partial S_{(t;x)}^{(m)}}\Omega ^{(m-1)}[\varphi ^{(0)}(x),\psi
^{(0)}(\mu )dx]-\int_{\partial S_{(t_{0};x_{0})}^{(m)}}\Omega
^{(m-1)}[\varphi ^{(0)}(\lambda ),\psi ^{(0)}(\mu )dx]  \notag \\
&=&\int_{S_{\pm }^{(m)}(\sigma _{(t;x)}^{(m-1)},\sigma
_{(t_{0};x_{0})}^{(m-1)})}d\Omega ^{(m-1)}[\varphi ^{(0)}(\lambda ),\psi
^{(0)}(\mu )dx]  \notag \\
&=&\int_{S_{\pm }^{(m)}(\sigma _{(t;x)}^{(m-1)},\sigma
_{(t_{0};x_{0})}^{(m-1)})}Z^{(m)}[\varphi ^{(0)}(\lambda ),\psi ^{(0)}(\mu
)dx],  \notag
\end{eqnarray}%
where, by definition, $m$-dimensional open surfaces $S_{\pm }^{(m)}(\sigma
_{(t;x)}^{(m-1)},\sigma _{(t_{0};x_{0})}^{(m-1)})\subset M_{\mathrm{T}}$ are
spanned smoothly without self-intersection between two homological cycles $%
\sigma _{(t;x)}^{(m-1)}=\partial S_{(t;x)}^{(m)}$ and $\sigma
_{(t_{0};x_{0})}^{(m-1)}=\partial S_{(t_{0};x_{0})}^{(m)}\in C_{m-1}(M_{%
\mathrm{T}};\mathbb{C})$ in such a way that the boundary $\partial
(S_{+}^{(m)}(\sigma _{(t_{0};x_{0})}^{(m-1)},\sigma
_{(t_{0};x_{0})}^{(m-1)})\cup S_{-}^{(m)}(\sigma _{(t;x)}^{(m-1)},\sigma
_{(t_{0};x_{0})}^{(m-1)}))=\oslash .$ Making use of the relationship (\ref%
{eq:2.15}), one can thereby find easily the following integral operator
expressions in $\mathcal{H}_{-}$:
\begin{eqnarray}
\mathbf{\Omega }_{\pm } &=&\mathbf{1}-\int_{\Sigma }d\rho (\eta )\tilde{\psi}%
^{(0)}(\xi )\Omega _{(t_{0};x_{0})}^{-1}(\xi ,\eta )  \label{eq:2.16} \\
&&\times \int_{S_{\pm }^{(m)}(\sigma _{(t;x)}^{(m-1)},\sigma
_{(t_{0};x_{0})}^{(m-1)})}Z^{(m)}[\varphi ^{{(0)}}(\eta ),(\cdot )dx]  \notag
\end{eqnarray}%
defined for fixed pairs $(\varphi ^{(0)}(\xi ),\psi ^{(0)}(\eta ))\in
\mathcal{H}_{0}^{\ast }\times \mathcal{H}_{0}$ and $(\tilde{\varphi}%
^{(0)}(\xi ),\tilde{\psi}^{(0)}(\mu ))\in \mathcal{\tilde{H}}_{0}^{\ast
}\times \mathcal{\tilde{H}}_{0},$ $\lambda ,\mu \in \Sigma ,$ being bounded
invertible operators of Volterra type \cite{GK,My,Bu,DS} on the whole
Hilbert space $\mathcal{H}.$ Moreover, for the differential operators $%
\mathrm{\tilde{L}}_{j}:\mathcal{H}\longrightarrow \mathcal{H},$ $\ j=%
\overline{1,2},$ one can get easily the following expressions
\begin{equation}
\tilde{\mathrm{L}}_{j}=\mathbf{\Omega }_{\pm }\mathrm{L}_{j}\mathbf{\Omega }%
_{\pm }^{-1}  \label{eq:2.17}
\end{equation}%
for $j=\overline{1,2},$ where the left hand-side of (\ref{eq:2.17}) does not
depend on sings "$\pm $" of the right-hand sides. Thereby, the Volterrian
integral operators (\ref{eq:2.16}) are the Delsarte-Darboux transmutation
operators, mapping a given set $\mathcal{L}$ of differential operators into
a new set $\mathcal{\tilde{L}}$ of differential operators transformed via
the Delsarte expressions (\ref{eq:2.17}).

\textbf{2.2 } Suppose now that all of differential operators $%
L_{j}(t;x|\partial ),$ $j=\overline{1,2},$ considered above don't depend one
the variable $t\in \mathrm{T.}$ Then, evidently, one can take
\begin{eqnarray}
\mathcal{H}_{0} &:&=\{\psi _{\mu }^{(0)}(\xi )\in L_{2.-}(M;\mathbb{C}%
^{N}):L_{j}\psi _{\mu }^{(0)}(\xi )=\mu _{j}\psi _{\mu }^{(0)}(\xi ),  \notag
\\
j &=&\overline{1,2},\text{ }\psi _{\mu }^{(0)}(\xi )|_{\tilde{\Gamma}}=0,\mu
=(\mu _{1},\mu _{2})\in \sigma (\tilde{L})\cap \overline{\sigma }(L^{\ast
}),\xi \in \Sigma _{\sigma }\}  \notag \\
\mathcal{\tilde{H}}_{0} &:&=\{\tilde{\psi}_{\mu }^{(0)}(\xi )\in L_{2.-}(M;%
\mathbb{C}^{N}):\tilde{L}_{j}\tilde{\psi}_{\mu }^{(0)}(\xi )=\mu _{j}\tilde{%
\psi}_{\mu }^{(0)}(\xi ),  \notag \\
j &=&\overline{1,2},\text{ }\tilde{\psi}_{\mu }^{(0)}(\xi )|_{\tilde{\Gamma}%
}=0,\mu =(\mu _{1},\mu _{2})\in \sigma (\tilde{L})\cap \overline{\sigma }%
(L^{\ast }),\xi \in \Sigma _{\sigma }\}  \notag \\
\mathcal{H}_{0}^{\ast } &:&=\{\varphi _{\lambda }^{(0)}(\eta )\in L_{2.-}(M;%
\mathbb{C}^{N}):L_{j}^{\ast }\varphi _{\lambda }^{(0)}(\eta )=\bar{\lambda}%
_{j}\varphi _{\lambda }^{(0)}(\eta ),j=\overline{1,2},  \label{eq:2.18} \\
\varphi _{\lambda }^{(0)}(\eta )|_{\tilde{\Gamma}} &=&0,\lambda =(\lambda
_{1},\lambda _{2})\in \sigma (\tilde{L})\cap \overline{\sigma }(L^{\ast
}),\eta \in \Sigma _{\sigma }\}  \notag \\
\mathcal{\tilde{H}}_{0}^{\ast } &:&=\{\tilde{\varphi}_{\lambda }^{(0)}(\eta
)\in L_{2.-}(M;\mathbb{C}^{N}):\tilde{L}_{j}^{\ast }\tilde{\varphi}_{\lambda
}^{(0)}(\eta )=\bar{\lambda}_{j}\varphi _{\lambda }^{(0)}(\eta ),  \notag \\
j &=&\overline{1,2},\text{ }\tilde{\varphi}_{\lambda }^{(0)}(\eta )|_{\tilde{%
\Gamma}}=0,\lambda =(\lambda _{1},\lambda _{2})\in \sigma (\tilde{L})\cap
\overline{\sigma }(L^{\ast }),\eta \in \Sigma _{\sigma }\}  \notag
\end{eqnarray}%
and construct the corresponding Delsarte-Darboux transmutation operators
\begin{eqnarray}
\mathbf{\Omega }_{\pm } &=&1-\int_{\sigma (\tilde{L})\cap \overline{\sigma }%
(L^{\ast })}d\rho _{\sigma }(\lambda )\int_{\Sigma _{\sigma }\times \Sigma
_{\sigma }}d\rho _{\Sigma _{\sigma }}(\xi )d\rho _{\Sigma _{\sigma }}(\eta )
\label{eq:2.19} \\
&&\times \int_{S_{\pm }^{(m)}\sigma _{(t_{0};x_{0})}^{(m-1)},\sigma
_{(t_{0};x_{0})}^{(m-1)}}dx\tilde{\psi}_{\lambda }^{(0)}(\xi )\Omega
_{x_{0}}^{-1}(\lambda ;\xi ;\eta )\bar{\varphi}_{\lambda }^{(0),\mathrm{%
\intercal }}(\eta )(\cdot )  \notag
\end{eqnarray}%
acting already in the Hilbert space $L_{2,+}(M;\mathbb{C}^{N}),$ where for
any $(\lambda ;\xi ,\eta )\in (\sigma (\tilde{L})\cap \overline{\sigma }%
(L^{\ast })\times \Sigma _{\sigma }^{2}$ kernels
\begin{equation}
\Omega _{(x_{0})}(\lambda ;\xi ,\eta ):=\int_{\sigma _{x_{0}}^{(m-1)}}\Omega
^{(m-1)}[\varphi _{\lambda }^{(0)}(\xi ),\psi _{\lambda }^{(0)}(\eta )dx]
\label{eq:2.20}
\end{equation}%
for $(\xi ,\eta )\in \Sigma _{\sigma }^{2}$ and every $\lambda \in \sigma (%
\tilde{L})\ \cap \overline{\sigma }(L^{\ast })$ belong to $L_{2}^{(\rho
)}(\Sigma _{\sigma };\mathbb{C})\otimes L_{2}^{(\rho )}(\Sigma _{\sigma };%
\mathbb{C}).$ Moreover, as $\partial \mathbf{\Omega }_{\pm }/\partial
t_{j}=0,$ $j=\overline{1,2,}$ \ one gets easily the set of differential
expressions
\begin{equation}
\tilde{L}_{j}(x|\partial ):=\mathbf{\Omega }_{\pm }L_{j}(x|\partial )\mathbf{%
\Omega }_{\pm }^{-1}  \label{eq:2.21}
\end{equation}%
$j=\overline{1,2},$ also commuting, evidently, to each other.

The Volterrian operators (\ref{eq:2.19}) possess some additional properties
Namely, define the following Fredholm type integral operator in $H:$
\begin{equation}
\mathbf{\Omega :=\Omega }_{+}^{-1}\mathbf{\Omega }_{-},  \label{eq:2.22}
\end{equation}%
which can be written in the form
\begin{equation}
\mathbf{\Omega }\mathbb{=}\mathbf{1}\mathbb{+}\Phi (\mathbf{\Omega })\mathbf{%
,}  \label{eq:2.23}
\end{equation}%
where the operator $\ \Phi (\mathbf{\Omega })\in \mathcal{B}_{\infty }(H)$
is compact. Moreover, due to the relationships (\ref{eq:2.21}) one gets
easily that the following commutator conditions
\begin{equation}
\lbrack \mathbf{\Omega },L_{j}]=0  \label{eq:2.24}
\end{equation}%
hold for $j=\overline{1,2}$.

Denote now by $\hat{\Phi}\mathbf{(\Omega )}\in H_{-}\otimes H_{-}$ and $\hat{%
K}_{+}(\mathbf{\Omega }),$ $\hat{K}_{-}(\mathbf{\Omega })\in H_{-}\otimes
H_{-}$ the kernels corresponding \cite{Be,BS} to operators $\Phi (\mathbf{%
\Omega })\in \mathcal{B}_{\infty }(H)$ and $\mathbf{\Omega }_{\pm }-\mathbf{1%
}\in \mathcal{B}_{\infty }(H)$. Then due to the fact that supports $supp$ $%
K_{+}(\mathbf{\Omega })\cap suppK_{-}(\mathbf{\Omega })=\sigma
_{x}^{(m-1)}\cup \sigma _{x_{0}}^{(m-1)},$ one gets from (\ref{eq:2.22}) and
(\ref{eq:2.23}) the well known Gelfand-Levitan-Marchenko linear integral
equation
\begin{equation}
\hat{K}_{+}(\mathbf{\Omega })+\hat{\Phi (\mathbf{\Omega })}+\hat{K}_{+}(%
\mathbf{\Omega })_{+}\ast \hat{\Phi}(\mathbf{\Omega })=\hat{K}_{-}(\mathbf{%
\Omega }),  \label{eq:2.25}
\end{equation}%
allowing to find the factorizing the Fredholmian operator (\ref{eq:2.22})
kernel $\hat{K}_{+}(\mathbf{\Omega })(x;y)\in H_{-}\otimes H_{-}$ \ \ for
all $\ y\in suppK_{+}(\mathbf{\Omega })$. The conditions (\ref{eq:2.24}) can
be rewritten suitably as follows:
\begin{equation}
(L_{j,ext}\otimes \mathbf{1})\hat{\Phi}(\mathbf{\Omega })=(1\otimes
L_{j,ext}^{\ast })\hat{\Phi}(\mathbf{\Omega })  \label{eq:2.26}
\end{equation}%
for $j=\overline{1,2},$ where $L_{j,ext}\in \mathcal{L}(H_{-}),$ $\ j=%
\overline{1,2},$ and their adjoint $L_{j,ext}^{\ast }\in \mathcal{L}(H_{-}),$
$\ j=\overline{1,2},$ are the corresponding extensions \cite{Be,PSPS,BS} of
the differential operators $L_{j}$ and $L_{j}^{\ast }\in \mathcal{L}(H),$ $j=%
\overline{1,2}.$

Concerning the relationships (\ref{eq:2.21}) one can write down \cite%
{Be,PSPS} kernel conditions similar to (\ref{eq:2.26}):
\begin{equation}
(\tilde{L}_{j,ext}\otimes \mathbf{1})\hat{K}_{\pm }(\mathbf{\Omega })=(%
\mathbf{1}\otimes L_{j,ext}^{\ast })\hat{K}_{\pm }(\mathbf{\Omega }),
\label{eq:2.27}
\end{equation}%
where as above, $\tilde{L}_{j,ext}\in \mathcal{L}(H_{-}),$ $j=\overline{1,2}$
are the corresponding rigging extensions of the differential operators $%
\tilde{L}_{j}\in \mathcal{L}(H),$ $j=\overline{1,2}$. \newline

\textbf{2.3 } Proceed now to analyzing the question about the general
differential and spectral structure of transformed operator expression (\ref%
{eq:2.17}). It is evidently that found above conditions (\ref{eq:2.25}) and (%
\ref{eq:2.26}) on the kernels $\hat{K}_{\pm }(\mathbf{\Omega })\in \mathcal{H%
}_{-}\otimes \mathcal{H}_{-}$ of Delsarte- Darboux transmutation operators
are necessary for the operator expressions (\ref{eq:2.17}) to exist and be
differential. Put the question whether these conditions are also sufficient?

For studying this question let us consider Volterrian operators \ (\ref%
{eq:2.16}) and (\ref{eq:2.19}) with kernels satisfying the conditions (\ref%
{eq:2.25}) and (\ref{eq:2.26}), assuming that suitable oriented surfaces $%
S_{\pm }^{(m)}(\sigma _{(t;x)^{(m-1)}},\sigma _{(t_{0};x_{0})^{(m-1)}})\in
C_{m}(M_{\mathrm{T}};\mathbb{C})$ are given as follows:
\begin{eqnarray}
S_{+}^{(m)}(\sigma _{(t;x)^{(m-1)}},\sigma _{(t_{0};x_{0})^{(m-1)}})
&=&\{(t^{\prime };x^{\prime })\in M_{\mathrm{T}}:  \notag \\
t^{\prime } &=&P(t;x|x^{\prime }),\text{ }t\in \mathrm{T}\},  \notag \\
S_{-}^{(m)}(\sigma _{(t;x)^{(m-1)}},\sigma _{(t_{0};x_{0})^{(m-1)}})
&=&\{(t^{\prime };x^{\prime })\in M_{\mathrm{T}}:  \notag \\
t^{\prime } &=&P(t;x|x^{\prime })\in \mathrm{T}\backslash \lbrack t_{0},t]\},
\label{eq:2.28}
\end{eqnarray}%
where a mapping $P\in C^{\infty }(M_{\mathrm{T}}\times M;\mathrm{T})$ is
smooth and such that the boundaries $\partial S_{\pm }^{(m)}(\sigma
_{(t;x)}^{(m-1)},\sigma _{(t_{0};x_{0})}^{(m-1)})$ $=\pm (\sigma
_{(t;x)}^{(m-1)}-\sigma _{(t_{0};x_{0})}^{(m-1)})$ with cycles $\sigma
_{(t;x)}^{(m-1)}$ and $\sigma _{(t_{0};x_{0})}^{(m-1)}\in \mathcal{K}(M_{%
\mathrm{T}})$ being homological to each other for any choice of points $%
(t_{0};x_{0})$ and $(t;x)\ \in M_{\mathrm{T}}.$ Then one can see by means of
some simple but cumbersome calculations, based on considerations from \cite%
{GS} and \cite{Fa}, that the resulting expressions on the right hand-sides
of
\begin{equation}
\mathrm{\tilde{L}}=\mathrm{L}+[\mathrm{K}_{\pm }(\mathbf{\Omega }),\mathrm{L}%
]\cdot \mathbf{\Omega }_{\pm }^{-1}  \label{eq:2.29}
\end{equation}%
are exactly equal to each other differential ones if such there was the
expression for an operator $\mathrm{L}\in \mathcal{L}(\mathcal{H}).$

Concerning the inverse operators $\mathbf{\Omega }_{\pm }^{-1}\in \mathcal{B}%
(\mathcal{H})$ present in (\ref{eq:2.29}) one can notice here that due to
the functional symmetry between closed subspaces $\mathcal{H}_{0}$ and $%
\mathcal{\tilde{H}}_{0}\subset \mathcal{\tilde{H}}_{-},$ the defining
relationships (\ref{eq:2.14}) and (\ref{eq:2.4}) are reversible, that is
there exist the inverse operator mappings $\mathbf{\Omega }_{\pm }^{-1}:%
\mathcal{\tilde{H}}_{0}\rightarrow \mathcal{H}_{0},$ such that
\begin{equation}
\mathbf{\Omega }_{\pm }^{-1}:\tilde{\psi}^{(0)}(\lambda )\longrightarrow
\psi ^{(0)}(\lambda ):=\tilde{\psi}^{(0)}(\lambda )\cdot \tilde{\Omega}%
_{(t;x)}^{-1}\tilde{\Omega}_{(t;x)}  \label{eq:2.30}
\end{equation}%
for some suitable kernels $\tilde{\Omega}_{(t;x)}(\lambda ,\mu )$ and $%
\tilde{\Omega}_{(t_{0};x_{0})}(\lambda ,\mu )\ \in L_{2}^{(\rho )}(\Sigma ;%
\mathbb{C})\otimes L_{2}^{(\rho )}(\Sigma ;\mathbb{C}),$ related naturally
with the transformed differential expression $\mathrm{\tilde{L}}\in \mathcal{%
L}(\mathcal{H}).$ Thereby, due to the expressions (\ref{eq:2.30}) one can
write down similar to (\ref{eq:2.19}) the following inverse integral
operators:
\begin{eqnarray}
\mathbf{\Omega }_{\pm }^{-1} &=&\mathbf{1}-\int_{\Sigma }d\rho (\xi
)\int_{\Sigma }d\rho (\eta )\psi ^{(0)}(\xi )\tilde{\Omega}%
_{t_{0};x_{0}}^{-1}(\xi ,\eta )  \label{eq:2.31} \\
&&\times \int_{S_{\pm }^{(m)}(\sigma _{(t;x)}^{(m-1)},\sigma
_{(t_{0};x_{0})}^{(m-1)})}\tilde{Z}^{(m)}[\tilde{\varphi}^{(0)}(\eta
),(\cdot )dx]  \notag
\end{eqnarray}%
defined for fixed pairs $(\tilde{\varphi}^{(0)}(\xi ),\tilde{\psi}%
^{(0)}(\eta ))\in \mathcal{\tilde{H}}_{0}^{\ast }\times \mathcal{\tilde{H}}%
_{0}$ and $(\varphi ^{(0)}(\xi ),\psi ^{(0)}(\eta ))\in \mathcal{H}%
_{0}^{\ast }\times \mathcal{H}_{0},$ $\xi ,\eta \in \Sigma ,$ and being
bounded invertible operators of Volterra type in the whole Hilbert space $%
\mathcal{H}.$ In particular, the compatibility conditions $\mathbf{\Omega }%
_{\pm }\mathbf{\Omega }_{\pm }^{-1}=\mathbf{1}=\mathbf{\Omega }_{\pm }^{-1}%
\mathbf{\Omega }_{\pm }$ must be fulfilled identically in $\mathcal{H},$
involving some restrictions identifying measures $\rho $ and $\Sigma $ and
possible asymptotic conditions of coefficient functions of the differential
expression $\mathrm{L}\in \mathcal{L}.$ Such kinds of restrictions were
already mentioned before in \cite{ZS,Ko,Za}, where in particular the
relationships with the local and nonlocal Riemann problems were discussed.

\textbf{2.4 } Within the framework of the general construction presented
above one can give a natural interpretation of so called Backlund
transformations for coefficient functions of a given differential operator
expression $\mathrm{L}\in \mathcal{L}(\mathcal{H})$. Namely, following the
symbolic considerations in \cite{LPS}, we reinterpret the approach devised
there for constructing the Backlund transformations making use of the
techniques based on the theory of Delsarte transmutation operators. Let us
define two different Delsarte-Darboux transformed differential operator
expressions
\begin{equation}
\mathrm{L}_{1}=\mathbf{\Omega }_{1,\pm }\mathrm{L}\mathbf{\Omega }_{1,\pm
}^{-1},\qquad \mathrm{L}_{2}=\mathbf{\Omega }_{2,\pm }\mathrm{L}\mathbf{%
\Omega }_{2,\pm }^{-1},  \label{eq:2.32}
\end{equation}%
where $\mathbf{\Omega }_{1,+},\mathbf{\Omega }_{2,-}\in \mathcal{B}(\mathcal{%
H})$ are some Delsarte transmutation Volterrian operators in $\mathcal{H}$
with Borel spectral measures $\rho _{1}$ and $\rho _{2}$ on $\Sigma ,$ such
that the following conditions
\begin{equation}
\mathbf{\Omega }_{1,+}^{-1}\mathbf{\Omega }_{1,-}=\mathbf{\Omega }=\mathbf{%
\Omega }_{2,+}^{-1}\mathbf{\Omega }_{2,-}  \label{eq:2.33}
\end{equation}%
hold. Making use now of the conditions (\ref{eq:2.32}) and relationships (%
\ref{eq:2.33}) one finds easily that the operator $\mathrm{B}:=\mathbf{%
\Omega }_{2,-}\mathbf{\Omega }_{1,+}^{-1}\in \mathcal{B}(\mathcal{H})$
satisfies the following operator equations:
\begin{equation}
\mathrm{L}_{2}\mathrm{B=BL}_{1},\quad \mathbf{\Omega }_{2,\pm }\mathrm{B=B}%
\mathbf{\Omega }_{1,\pm },  \label{eq:2.34}
\end{equation}%
which motivate the next definition.

\begin{definition}
An invertible symbolic mapping $\mathrm{B}:\mathcal{L}(\mathcal{H}%
)\longrightarrow \mathcal{L}$ $\mathcal{(H)}$ \ will be called a
Darboux-Backlund transformation of an operator $\mathrm{L}_{1}\in \mathcal{L}%
(\mathcal{H})$\ into the operator $\mathrm{L}_{2}\in \mathcal{L}(\mathcal{H}%
) $\ if there holds the condition
\begin{equation}
\lbrack \mathrm{QB},\mathrm{L}_{1}]=0  \label{eq:2.35}
\end{equation}%
for some linear differential expression $\mathrm{Q}\in \mathcal{L}(\mathcal{H%
}).$
\end{definition}

The condition (\ref{eq:2.35}) can be realized as follows. Take any
differential expression $\mathrm{q}\in \mathcal{L}(\mathcal{H})$
satisfying the symbolic equation
\begin{equation}
\lbrack \mathrm{qB},\mathrm{L}]=0.  \label{eq:2.36}
\end{equation}%
Then, making use of the transformations like (\ref{eq:2.32}), from (\ref%
{eq:2.33}) one finds that
\begin{equation}
\lbrack \mathrm{QB},\mathrm{L}_{1}]=0,  \label{eq:2.37}
\end{equation}%
where owing to (\ref{eq:2.34})
\begin{equation}
\mathrm{QB}:=\mathbf{\Omega }_{1,+}q\mathrm{B}\mathbf{\Omega }_{1,+}^{-1}=%
\mathbf{\Omega }_{1,+}\mathrm{q}\mathbf{\Omega }_{2,+}^{-1}\mathrm{B}.
\label{eq:2.38}
\end{equation}%
Therefore, the expression $\mathrm{Q}=\mathbf{\Omega }_{1,+}q\mathbf{\Omega }%
_{2,+}^{-1}$ appears to be differential one too owing to the conditions (\ref%
{eq:2.34}).

The consideration above related with the symbolic mapping $\mathrm{B:}%
\mathcal{L(H)\rightarrow L(H)}$ gives rise to an effective tool of
constructing self-Backlund transformations for coefficients of differential
operator expressions $\mathrm{L}_{1},\mathrm{L}_{2}\in \mathcal{L(H)}$
having many applications \cite{MS,No,PM,DS,SP} \ in \ spectral and soliton
theories.

\textbf{2.5 \ }Return now back to studying the structure Delsarte-Darboux
transformations for a polynomial differential operators pencil
\begin{equation}
\mathrm{L}(\lambda ;x|\partial ):=\sum_{j=0}^{n(L)}L_{j}(x|\partial )\lambda
^{j},  \label{2.39}
\end{equation}%
where $n(L)\in \mathbb{Z}_{+}$ and $\lambda \in \mathbb{C}$ is a
complex-valued parameter. It is asked to find the corresponding to (\ref%
{2.39}) Delsarte-Darboux transformations $\mathbf{\Omega }_{\lambda ,\pm
}\in \mathcal{B(H)},$ $\lambda \in \mathbb{C},$ such that for some
polynomial differential operators pencil $\mathrm{\tilde{L}}(\lambda
;x|\partial )\in \mathcal{L(H)}$ the following Delsarte-Lions \cite{DL}
\bigskip transmutation condition
\begin{equation}
\mathrm{\tilde{L}}\mathbf{\Omega }_{\lambda ,\pm }=\mathbf{\Omega }_{\lambda
,\pm }\mathrm{L}  \label{2.40}
\end{equation}%
holds for almost all $\lambda \in \mathbb{C}.$ For such transformations \ $%
\mathbf{\Omega }_{\lambda \pm }\in \mathcal{B(H)}$ to be found, let us
consider a parameter $\tau \in \mathbb{R}$ dependent differential operator $%
\mathrm{L}_{\tau }(x|\partial )\in \mathcal{L(H}_{\tau }\mathcal{)},$ where
\begin{equation}
\mathrm{L}_{\tau }(x|\partial ):=\sum_{j=0}^{n(L)}L_{j}(x|\partial )\partial
^{j}/\partial \tau ^{j},  \label{2.41}
\end{equation}%
acting in the functional space $\mathcal{H}_{\tau }=C^{q(L)}(\mathbb{R}%
_{\tau };\mathcal{H})$ for some $q(L)\in \mathbb{Z}_{+}.$ Then one can
easily construct the corresponding Delsarte-Darboux transformations $\mathbf{%
\Omega }_{\tau ,\pm }\in \mathcal{B(H}_{\tau }\mathcal{)}$ of Volterra type
for some differential operator expression%
\begin{equation}
\mathrm{\tilde{L}}_{\tau }(x|\partial ):=\sum_{j=0}^{n(L)}\tilde{L}%
_{j}(x|\partial )\partial ^{j}/\partial \tau ^{j},  \label{2.42}
\end{equation}%
if the following Delsarte-Lions \cite{DL} transmutation conditions%
\begin{equation}
\mathrm{\tilde{L}}_{\tau }\mathbf{\Omega }_{\tau ,\pm }=\mathbf{\Omega }%
_{\tau ,\pm }\mathrm{L}_{\tau }  \label{2.43}
\end{equation}%
hold in $\mathcal{H}_{\tau }.$ Thus, making use of the results obtained
above, one can write down that
\begin{eqnarray}
\mathbf{\Omega }_{\tau ,\pm } &=&\mathbf{1}-\int_{\Sigma }d\rho _{\Sigma
}(\xi )\int_{\Sigma }d\rho _{\Sigma }(\eta )\tilde{\psi}_{\tau
}^{(0)}(\lambda ;\xi )\Omega _{(\tau _{0};x_{0})}^{-1}(\lambda ;\xi ,\eta )
\label{2.44} \\
&&\times \int_{S_{\pm }^{(m)}(\sigma _{(\tau ;x)}^{(m-1)},\sigma _{(\tau
_{0};x_{0})}^{(m-1)})}Z^{(m)}[\varphi _{\tau }^{(0)}(\lambda ;\eta ),(\cdot
)dx]  \notag
\end{eqnarray}%
defined by means of the following closed subspaces $\mathcal{H}_{\tau
,0}\subset \mathcal{H}_{\tau ,-}$ and \bigskip $\mathcal{H}_{\tau ,0}^{\ast
}\subset \mathcal{H}_{\tau ,-}^{\ast }:$%
\begin{eqnarray*}
\mathcal{H}_{\tau ,0} &:&=\{\psi _{\tau }^{(0)}(\lambda ;\xi )\in \mathcal{H}%
_{\tau ,-}:\mathrm{L}_{\tau }\psi _{\tau }^{(0)}(\lambda ;\xi )=0,\text{ } \\
\psi _{\tau }^{(0)}(\lambda ;\xi )|_{\tau =0} &=&\psi ^{(0)}(\lambda ;\xi
)\in \mathcal{H},\text{ }\mathrm{L}\psi ^{(0)}(\lambda ;\xi )=0,\text{ } \\
\psi ^{(0)}(\lambda ;\xi )|_{\Gamma } &=&0,\text{ }\lambda \in \mathbb{C},%
\text{ }\xi \in \Sigma \},
\end{eqnarray*}

\bigskip
\begin{eqnarray}
\mathcal{H}_{\tau ,0}^{\ast } &:&=\{\varphi _{\tau }^{(0)}(\lambda ;\eta
)\in \mathcal{H}_{\tau ,-}^{\ast }:\mathrm{L}_{\tau }\varphi _{\tau
}^{(0)}(\lambda ;\eta )=0,\text{ }  \label{2.45} \\
\varphi _{\tau }^{(0)}(\lambda ;\eta )|_{\tau =0} &=&\varphi ^{(0)}(\lambda
;\eta )\in \mathcal{H}^{\ast },\text{ }\mathrm{L}\varphi ^{(0)}(\lambda
;\eta )=0,\text{ }  \notag \\
\varphi ^{(0)}(\lambda ;\eta )|_{\Gamma } &=&0,\text{ }\lambda \in \mathbb{C}%
,\text{ }\eta \in \Sigma \}.  \notag
\end{eqnarray}%
Recalling now that our operators $L_{j}\in \mathcal{L(H)},$ $j=\overline{%
0,r(L)},$ do not depend on the parameter $\tau \in \mathbb{R},$ one can
derive easily from (\ref{2.44})
\begin{eqnarray}
\mathbf{\Omega }_{\pm } &=&\mathbf{1}-\int_{\Sigma }d\rho _{\Sigma }(\xi
)\int_{\Sigma }d\rho _{\Sigma }(\eta )\tilde{\psi}^{(0)}(\lambda ;\xi
)\Omega _{(x_{0})}^{-1}(\lambda ;\xi ,\eta )  \label{2.46} \\
&&\times \int_{S_{\pm }^{(m)}(\sigma _{(x)}^{(m-1)},\sigma
_{(x_{0})}^{(m-1)})}Z_{0}^{(m)}[\varphi ^{(0)}(\lambda ;\eta ),(\cdot )dx],
\notag
\end{eqnarray}%
where we put $\sigma _{x}^{(m-1)}:=\sigma _{(\tau _{0};x)}^{(m-1)},$ $\sigma
_{x_{0}}^{(m-1)}:=\sigma _{(\tau _{0};x_{0})}^{(m-1)}\in C_{m-1}(\mathbb{R}%
^{m};\mathbb{C})$ and
\begin{equation}
Z_{0}^{(m)}[\varphi ^{(0)}(\lambda ;\eta ),\psi ^{(0)}dx]:=Z^{(m)}[\varphi
_{\tau }^{(0)}(\lambda ;\eta ),\psi _{\tau }^{(0)}dx]|_{d\tau =0}.
\label{2.47}
\end{equation}%
\bigskip The corresponding to (\ref{2.46}) closed subspaces $\mathcal{H}%
_{0}\in \mathcal{H}_{-}$ and $\mathcal{H}_{0}^{\ast }\in \mathcal{H}%
_{-}^{\ast }$ are given as follows:\bigskip
\begin{equation}
\mathcal{H}_{0}:=\{\psi ^{(0)}(\lambda ;\xi )\in \mathcal{H}_{-}:\mathrm{L}%
\psi ^{(0)}(\lambda ;\xi )=0,\text{ }\psi ^{(0)}(\lambda ;\xi )|_{\Gamma }=0,%
\text{ }\lambda \in \mathbb{C},\text{ }\xi \in \Sigma \},  \label{2.48}
\end{equation}%
\begin{equation}
\mathcal{H}_{\tau ,0}^{\ast }:=\{\varphi ^{(0)}(\lambda ;\eta )\in \mathcal{H%
}_{-}^{\ast }:\mathrm{L}\varphi ^{(0)}(\lambda ;\eta )=0,\text{ }\varphi
^{(0)}(\lambda ;\eta )|_{\Gamma }=0,\text{ }\lambda \in \mathbb{C},\text{ }%
\eta \in \Sigma \}.  \notag
\end{equation}%
Thereby, making use of the expressions (\ref{2.46}) one can construct the
Delsarte-Darboux transformed linear differential pencil $\mathrm{\tilde{L}}%
\in \mathcal{L(H)},$ whose coefficients are related with those of the pencil
$\mathrm{L}\in \mathcal{L(H)}$ via some Backlund type relationships useful
for applications \ (see \cite{SP,SPS,We,Gu,Ko}) in the soliton theory.

\section{Delsarte-Darboux type transmutation operators for special
multi-dimensional expressions and their applications}

\textbf{3.1} \textbf{A perturbed self-adjoint Laplace operator in }$\mathbb{R%
}^{n}.$\bigskip\ Consider the Laplace operator $-\Delta _{m}$ in $H:=L(%
\mathbb{R}^{m};\mathbb{C})$ perturbed by the multiplication operator on a
function $q\in W_{2}^{2}(\mathbb{R}^{m};\mathbb{C}),$ that is the operator%
\begin{equation}
L(x|\partial ):=-\Delta _{m}+q(x),  \label{3.1}
\end{equation}%
where $x\in \mathbb{R}^{m}.$ The operator (\ref{3.1}) is self-adjoint in $H.$
Applying the results from the Section 1 to the differential expression (\ref%
{3.1}) in the Hilbert space $H,$ one can write down the following invertible
Delsarte-Darboux transmutation operators:%
\begin{eqnarray}
\mathbf{\Omega }_{\pm } &=&\mathbf{1}-\int_{\sigma (L)}d\rho _{\sigma }(\xi
)\int_{\sigma (L)}d\rho _{\sigma }(\xi )\int_{\Sigma _{\sigma }}d\rho
_{\Sigma _{\sigma }}(\xi )\int_{\Sigma _{\sigma }}d\rho _{\Sigma _{\sigma
}}(\eta )  \label{3.2} \\
&&\times \tilde{\psi}^{(0)}(\lambda ;\xi )\Omega _{(x_{0})}^{-1}(\lambda
;\xi ,\eta )\int_{S_{\pm }^{(m)}(\sigma _{(x)}^{(m-1)},\sigma
_{(x_{0})}^{(m-1)})}^{(0)}dy\bar{\varphi}^{(0)\intercal }(\lambda ;\eta
),(\cdot ),  \notag
\end{eqnarray}%
where $\sigma _{x}^{(m-1)}\in \mathcal{K}(\mathbb{R}^{m})$ is some closed \
maybe non-compact simplicial hyper-surface in $\mathbb{R}^{m}$ parametrized
\bigskip by a running point $x\in \sigma _{x}^{(m-1)},$ and $\sigma
_{x_{0}}^{(m-1)}\in \mathcal{K}\mathbb{(R}^{m}\mathbb{)}$ is a suitable
homological to $\sigma _{x}^{(m-1)}$ simplicial hypersurface in $\mathbb{R}%
^{m}$ parametrized by a point $x_{0}\in \sigma _{x_{0}}^{(m-1)}.$ There
exist \ exactly two $m$-dimensional subspaces spanning them, say $S_{\pm
}^{(m)}(\sigma _{x}^{(m-1)},\sigma _{x_{0}}^{(m-1)})\in \mathcal{K}\mathbb{(R%
}^{m}\mathbb{)},$ such that \ $S_{+}^{(m)}(\sigma _{x}^{(m-1)},\sigma
_{x_{0}}^{(m-1)})\cup $\ $S_{-}^{(m)}(\sigma _{x}^{(m-1)},\sigma
_{x_{0}}^{(m-1)})=\mathbb{R}^{m}.$\ Taking into account these subspaces, one
can rewrite down compactly the Delsarte-Darboux transmutation operators (\ref%
{3.2}) for (\ref{3.1}):%
\begin{equation}
\mathbf{\Omega }_{\pm }=\mathbf{1+}\int_{S_{\pm }^{(m)}(\sigma
_{x}^{(m-1)},\sigma _{x_{0}}^{(m-1)})}dy\hat{K}_{\pm }(\mathbf{\Omega }%
)(x;y)(\cdot ),  \label{3.3}
\end{equation}%
where, as before, $x\in \sigma _{x}^{(m-1)}$ and kernels $\hat{K}_{\pm }(%
\mathbf{\Omega })\in H_{-}\otimes H_{-}$ satisfy the equations (\ref{eq:2.27}%
), or equivalently,
\begin{eqnarray}
&&-\Delta _{m}(x;\partial )\hat{K}_{\pm }(\mathbf{\Omega })(x;y)+\Delta
_{m}(y;\partial )\hat{K}_{\pm }(\mathbf{\Omega })(x;y)  \label{3.4} \\
&=&(q(y)-\tilde{q}(x))\hat{K}_{\pm }(\mathbf{\Omega })(x;y)  \notag
\end{eqnarray}%
\bigskip for all $x,y\in \sup $p$\hat{K}_{\pm }(\mathbf{\Omega }).$ Take for
simplicity, a non-compact closed simplicial hypersurface $\sigma
_{x}^{(m-1)}=$ $\sigma _{x,\gamma }^{(m-1)}:=\{y\in \mathbb{R}^{m}:<x-y,\pm
\gamma >=0\}$ and the degenerate simplicial cycle $\sigma
_{x_{0}}^{(m-1)}:=x_{0}=\infty \in \mathbb{R}^{m},$ where $\gamma \in
\mathbb{S}^{m-1}$ is an arbitrary versor, $||\gamma ||=0.$ Then, evidently,
\begin{equation}
S_{\pm }^{(m)}(\sigma _{x,\gamma )}^{(m-1)},\sigma _{\infty
}^{(m-1)}):=S_{\pm \gamma ,x}^{(m)}=\{y\in \mathbb{R}^{m}:<x-y,\pm \gamma >%
\text{ }\geq 0\}  \label{3.5}
\end{equation}%
and our transmutation operators (\ref{3.3}) take the form%
\begin{equation}
\mathbf{\Omega }_{\pm \gamma }=\mathbf{1+}\int_{S_{\pm \gamma ,x}^{(m)}}dy%
\hat{K}_{\pm \gamma }(\mathbf{\Omega })(x;y)(\cdot ),  \label{3.6}
\end{equation}%
where $\sup $p$\hat{K}_{\pm \gamma }(\mathbf{\Omega })=S_{\pm \gamma
,x}^{(m)},$ $S_{+\gamma ,x}^{(m)}\cap S_{-\gamma ,x}^{(m)}=\sigma _{x,\gamma
}^{(m-1)}\cup \sigma _{\infty }^{(m-1)}$ and $S_{+\gamma ,x}^{(m)}\cap
S_{-\gamma ,x}^{(m)}=\mathbb{R}^{m}$ for any direction $\gamma \in \mathbb{S}%
^{m-1}.$

The invertible transmutation Volterrian operators like (\ref{3.6}) were
constructed before by L.D. Faddeev \cite{Fa} for the self-adjoint perturbed
Laplace operator (\ref{3.1}) in $\mathbb{R}^{3}.$ He called them \cite{Fa}
transformation operators with a Volterrian direction $\gamma \in \mathbb{S}%
^{m-1}.$ It is easy to see that Faddeev's expressions (\ref{3.6}) are very
special cases of the general expressions (\ref{3.3}) obtained above.

Define now making use of (\ref{3.3}) the following Fredholmian operator in
the Hilbert space $H:$%
\begin{equation}
\mathbf{\Omega :=(1+}K_{+}\mathbf{(\Omega ))}^{-1}\mathbf{(1+}K_{-}\mathbf{%
(\Omega ))=1+}\Phi \mathbf{(\Omega )}  \label{3.7}
\end{equation}%
with the compact part $\Phi \mathbf{(\Omega )\in }\mathcal{B}_{\infty }%
\mathbf{(}H\mathbf{)}.$ Then the commutation equality
\begin{equation}
\lbrack L,\Phi \mathbf{(\Omega )]=0}  \label{3.8}
\end{equation}%
together with the Gelfand-Levitan-Marchenko equation
\begin{equation}
K_{+}\mathbf{(\Omega )+}\hat{\Phi}\mathbf{(\Omega )+}\hat{K}_{+}\mathbf{%
(\Omega )+}\hat{\Phi}\mathbf{(\Omega )=}\hat{K}_{-}\mathbf{(\Omega )}
\label{3.9}
\end{equation}%
for the corresponding kernels $\hat{K}_{\pm }\mathbf{(\Omega )}$ and $\hat{%
\Phi}\mathbf{(\Omega )\in }H_{-}\otimes H_{-}$ hold.

In \cite{Fa} there was thoroughly analyzed the spectral structure of kernels
$\hat{K}_{\pm }\mathbf{(\Omega )\in }H_{-}\otimes H_{-}$ in (\ref{3.6})
making use of the analytical properties of the corresponding Green functions
of the operator (\ref{3.1}). As one can see from (\ref{3.2}), these
properties depend strongly both on the structure of the spectral measures $%
\rho _{\sigma }$ on $\sigma (L)$ and $\rho _{\Sigma _{\sigma }}$ on $\Sigma
_{\sigma }$ and on analytical behavior of the kernel $\Omega _{\infty
}(\lambda ;\xi ,\eta )\in L_{2}^{(\rho )}($ $\Sigma _{\sigma };\mathbb{C}%
)\otimes L_{2}^{(\rho )}($ $\Sigma _{\sigma };\mathbb{C}),$ $\xi ,\eta \in
\Sigma _{\sigma },$ for \ all $\lambda \in \sigma (L).$ In \cite{Fa} there
was stated for any direction $\gamma \in \mathbb{S}^{m-1}$ the dependence of
kernels $\hat{K}_{\pm }\mathbf{(\Omega )\in }H_{-}\otimes H_{-}$ on the
regularized determinant of the resolvent $R_{\mu }(L)\in \mathcal{B}(H),$ $%
\mu \in \mathbb{C}/\sigma (L)$ is a regular point, for the operator (\ref%
{3.1}). This dependence can be also clarified if to make use of the approach
from Section 2.

\textbf{2.2 A two-dimensional Dirac type operator. }Let us define in $%
H:=L_{2}(\mathbb{R}^{2};\mathbb{C}^{2})$ a two-dimensional\bigskip\ Dirac
type operator
\begin{equation}
\mathrm{\tilde{L}}(x;\partial ):=\left(
\begin{array}{cc}
\partial /\partial x_{1} & \tilde{u}_{1}(x) \\
\tilde{u}_{2}(x) & \partial /\partial x_{2}%
\end{array}%
\right) ,  \label{3.10}
\end{equation}%
where $x:=(x_{1},x_{2})\in \mathbb{R}^{2},$ and coefficients $\tilde{u}%
_{j}\in W_{2}^{1}($ $\mathbb{R}^{2};\mathbb{C}),$ $j=\overline{1,2}.$ the
transformation properties of the operator \ (\ref{3.10}) were studied \cite%
{Ni} \ thoroughly by L.P. Nizhnik. In particular, he constructed some
special class of the Delsarte-Darboux transmutation operators in the form%
\begin{equation}
\mathbf{\Omega }_{\pm }=\mathbf{1+}\int_{S_{\pm }^{(2)}(\sigma
_{x}^{(1)},\sigma _{\infty }^{(1)})}dy\hat{K}_{\pm }(\mathbf{\Omega }%
)(x;y)(\cdot ),  \label{3.11}
\end{equation}%
where for two orthonormal versors $\gamma _{1}$ and $\gamma _{2}\in \mathbb{S%
}^{1},$ $||\gamma _{1}||=1=||\gamma _{2}||,$%
\begin{eqnarray}
S_{+}^{(2)}(\sigma _{x}^{(1)},\sigma _{\infty }^{(1)}) &:&=\{y\in \mathbb{R}%
^{2}:<x-y,\gamma _{1}>\geq 0\}  \label{3.12} \\
\cap \{y &\in &\mathbb{R}^{2}:<x-y,\gamma _{2}>\geq 0\},  \notag \\
S_{-}^{(2)}(\sigma _{x}^{(1)},\sigma _{\infty }^{(1)}) &:&=\{y\in \mathbb{R}%
^{2}:<x-y,\gamma _{1}>\leq 0\}  \notag \\
\cup \{y &\in &\mathbb{R}^{2}:<x-y,\gamma _{2}>\leq 0\}.  \notag
\end{eqnarray}%
In the case when $<x,\gamma _{j}>=x_{j}\in \mathbb{R},$ $j=\overline{1,2},$
the corresponding kernel
\begin{equation}
\hat{K}_{+}\mathbf{(\Omega )=}\left(
\begin{array}{cc}
K_{+,11}^{(1)}\delta _{<y-x,\gamma _{1}>}+K_{+,11}^{(0)}(x;y) &
K_{+,12}^{(1)}\delta _{<y-x,\gamma _{2}>}+K_{+,12}^{(0)}) \\
K_{+,21}^{(1)}\delta _{<y-x,\gamma _{1}>}+K_{+,21}^{(0)}(x;y) &
K_{+,22}^{(1)}\delta _{<y-x,\gamma _{2}>}+K_{+,22}^{(0)}%
\end{array}%
\right)  \label{3.13}
\end{equation}%
is Dirac delta-function singular, being, in part, localized on half-lines $%
<y-x,\gamma _{2}>=0$ and $<y-x,\gamma _{1}>=0,$ with regular all
coefficients $K_{+,ij}^{(l)}\in C^{1}(\mathbb{R}^{2}\times \mathbb{R}^{2};%
\mathbb{C})$ for all $i,j=\overline{1,2}$ and $l=\overline{0,1}.$ Such a
property of the transmutation kernels for the perturbed Laplace operator (%
\ref{3.1}) was also observed in \cite{Fa}, where it was motivated by the
necessary condition for the transformed operator $\tilde{L}(x;\partial )\in
\mathcal{L}(H)$\bigskip\ to be differential. As one can check, the same
reason of the existence of singularities holds in (\ref{3.13}).

Let us now consider the general expression like (\ref{3.3}) for the
corresponding hyper-surfaces $S_{\pm }^{(2)}(\sigma _{x}^{(1)},\sigma
_{\infty }^{(1)})$ spanning between a closed non-compact smooth cycle $%
\sigma _{x}^{(1)}$ $\in \mathcal{K}(\mathbb{R}^{2})$ and the infinite point $%
\sigma _{\infty }^{(1)}:=\infty \in \mathcal{K}(\mathbb{R}^{2}).$ A running
point $x\in \sigma _{x}^{(1)}$ is taken arbitrary but, as usual,
fixed\bigskip . the kernels $\hat{K}_{\pm }(\mathbf{\Omega })\in H_{-}\times
H_{-}$ in (\ref{3.11}) satisfy the standard conditions (\ref{eq:2.26}) and (%
\ref{eq:2.27}), that is

\bigskip
\begin{eqnarray}
(\mathrm{\tilde{L}}_{1,ext}\otimes \mathbf{1})\hat{K}_{\pm }(\mathbf{\Omega }%
) &=&(\mathbf{1}\otimes \mathrm{L}_{1,ext}^{\ast })\hat{K}_{\pm }(\mathbf{%
\Omega }),  \label{3.14} \\
\lbrack \mathrm{L}_{1},\Phi (\mathbf{\Omega })] &=&0  \notag
\end{eqnarray}%
for some matrix differential Dirac type operator $\mathrm{L}_{1}\in \mathcal{%
L}(H)$ of the form (\ref{3.1})$.$ Together with this Dirac operator the
following matrix second order differential operator%
\begin{equation}
\mathrm{\tilde{L}}_{2}:=\mathbf{1}\frac{\partial }{\partial t}+\left(
\begin{array}{cc}
\frac{\partial ^{2}}{\partial x_{1}^{2}}\pm \frac{\partial ^{2}}{\partial
x_{2}^{2}}-\tilde{v}_{2} & -2\frac{\partial \tilde{u}_{1}}{\partial x_{2}}
\\
-2\frac{\partial \tilde{u}_{2}}{\partial x_{1}} & \frac{\partial ^{2}}{%
\partial x_{1}^{2}}\pm \frac{\partial ^{2}}{\partial x_{2}^{2}}-\tilde{v}_{1}%
\end{array}%
\right)  \label{3.15}
\end{equation}%
in the parametric space $\mathcal{H}:=C^{1}(\mathbb{R};H)$\bigskip\ was
studied in \cite{Ni,NP}\ for which there was developed scattering theory and
given its application for constructing soliton-like exact solutions to the
so called Davey-Stewartson nonlinear dynamical system in partial
derivatives. The latter was based on the fact that two operators $\mathrm{%
\tilde{L}}_{1}$ and $\mathrm{\tilde{L}}_{2}\in \mathcal{L}(H)$ are commuting
to each other.

Namely, consider the Volterrian operators $\mathbf{\Omega }_{\pm }\in
\mathcal{B}(\mathcal{H})$ realizing the following Delsarte-Darboux
transmutations:\bigskip
\begin{equation}
\mathrm{\tilde{L}}_{1}\mathbf{\Omega }_{\pm }=\mathbf{\Omega }_{\pm }\mathrm{%
L}_{1},\text{ \ \ }\mathrm{\tilde{L}}_{2}\mathbf{\Omega }_{\pm }=\mathbf{%
\Omega }_{\pm }\mathrm{L}_{2}.  \label{3.16}
\end{equation}%
Here we put
\begin{eqnarray}
\mathrm{L}_{1}(x;\partial ) &:&=\left(
\begin{array}{cc}
\partial /\partial x_{1} & 0 \\
0 & \partial /\partial x_{2}%
\end{array}%
\right) ,  \label{3.17} \\
\mathrm{L}_{2} &:&=\mathbf{1}\frac{\partial }{\partial t}+\left(
\begin{array}{cc}
\frac{\partial ^{2}}{\partial x_{1}^{2}}\pm \frac{\partial ^{2}}{\partial
x_{2}^{2}}-\alpha _{2}(x_{2}) & 0 \\
0 & \frac{\partial ^{2}}{\partial x_{1}^{2}}\pm \frac{\partial ^{2}}{%
\partial x_{2}^{2}}-\alpha _{1}(x_{1})%
\end{array}%
\right) ,  \notag
\end{eqnarray}%
where $\alpha _{j}\in W_{2}^{1}(\mathbb{R};\mathbb{C}),$ $j=\overline{1,2},$
are some given functions. It is evident that operators (\ref{3.17}) are
commuting to each other. then, if the operators $\mathbf{\Omega }_{\pm }\in
\mathcal{B(}H\mathcal{)}$ exist and satisfy \bigskip (\ref{3.16}), the
following commutation condition
\begin{equation}
\lbrack \mathrm{\tilde{L}}_{1},\mathrm{\tilde{L}}_{2}]=0  \label{3.18}
\end{equation}%
holds, that there was exactly claimed above and effectively exploited before
in \cite{Ni,NP}.

Recall now that for the operators $\mathbf{\Omega }_{\pm }\in \mathcal{B(}H%
\mathcal{)}$ to exist they must satisfy additionally the kernel conditions (%
\ref{3.14}) and
\begin{eqnarray}
(\mathrm{\tilde{L}}_{2,ext}\otimes \mathbf{1})\hat{K}_{\pm }(\mathbf{\Omega }%
) &=&(\mathbf{1}\otimes \mathrm{L}_{2,ext}^{\ast })\hat{K}_{\pm }(\mathbf{%
\Omega }),  \label{3.19} \\
\lbrack \mathrm{L}_{2},\Phi (\mathbf{\Omega })] &=&0,  \notag
\end{eqnarray}%
where, as before, the operator $\Phi (\mathbf{\Omega )\in }\mathcal{B}%
_{\infty }\mathcal{(}H\mathcal{)}$ is defined by (\ref{3.7}) as
\begin{equation}
\mathbf{\Omega :=1+}\Phi (\mathbf{\Omega }).  \label{3.20}
\end{equation}%
Owing to the evident commutation condition (\ref{3.18}) the set of equations
(\ref{3.14}) and (\ref{3.19}) is compatible giving rise to the expression
like (\ref{3.11}), where the kernel $\hat{K}_{+}(\mathbf{\Omega })\in
H_{-}\otimes H_{-}$ satisfies the set of differential equations generalizing
those from \cite{Ni,NP}:%
\begin{eqnarray}
\frac{\partial K_{+,11}}{\partial x_{1}}+\frac{\partial K_{+,11}}{\partial
y_{1}}+\tilde{u}_{1}K_{+,21} &=&0,\text{ \ }\frac{\partial K_{+,12}}{%
\partial x_{1}}+\frac{\partial K_{+,12}}{\partial y_{1}}+\tilde{u}%
_{1}K_{+,22}=0,  \label{3.21} \\
\frac{\partial K_{+,21}}{\partial x_{2}}+\frac{\partial K_{+,21}}{\partial
x_{1}}+\tilde{u}_{2}K_{+,11} &=&0,\text{ \ }\frac{\partial K_{+,22}}{%
\partial x_{2}}+\frac{\partial K_{+,22}}{\partial y_{2}}+\tilde{u}%
_{2}K_{+,12}=0,  \notag
\end{eqnarray}%
\begin{eqnarray*}
\pm \frac{\partial \tilde{u}_{1}}{\partial x_{2}}K_{+,21} &=&\frac{\partial
K_{+,11}}{\partial t}+[(\frac{\partial ^{2}}{\partial x_{1}^{2}}-\frac{%
\partial ^{2}}{\partial y_{1}^{2}})\pm (\frac{\partial ^{2}}{\partial
x_{2}^{2}}-\frac{\partial ^{2}}{\partial y_{2}^{2}})]K_{+,11} \\
&&+(\alpha _{2}(x_{2})-\tilde{v}_{2}(x))K_{+,11} \\
\pm \frac{\partial \tilde{u}_{1}}{\partial x_{2}}K_{+,21} &=&\frac{\partial
K_{+,22}}{\partial t}+[(\frac{\partial ^{2}}{\partial x_{1}^{2}}-\frac{%
\partial ^{2}}{\partial y_{1}^{2}})\pm (\frac{\partial ^{2}}{\partial
x_{2}^{2}}-\frac{\partial ^{2}}{\partial y_{2}^{2}})]K_{+,22} \\
&&+(\alpha _{1}(x_{1})-\tilde{v}_{1}(x))K_{+,22}, \\
\mp 2\frac{\partial \tilde{u}_{1}}{\partial x_{2}}K_{+,22} &=&\frac{\partial
K_{+,12}}{\partial t}+[(\frac{\partial ^{2}}{\partial x_{1}^{2}}-\frac{%
\partial ^{2}}{\partial y_{1}^{2}})\pm (\frac{\partial ^{2}}{\partial
x_{2}^{2}}-\frac{\partial ^{2}}{\partial y_{2}^{2}})]K_{+,12} \\
&&+(\alpha _{1}(x_{1})-\tilde{v}_{2}(x))K_{+,22}. \\
2\frac{\partial \tilde{u}_{2}}{\partial x_{1}}K_{+,22} &=&\frac{\partial
K_{+,21}}{\partial t}+[(\frac{\partial ^{2}}{\partial x_{1}^{2}}-\frac{%
\partial ^{2}}{\partial y_{1}^{2}})\pm (\frac{\partial ^{2}}{\partial
x_{2}^{2}}-\frac{\partial ^{2}}{\partial y_{2}^{2}})]K_{+,21} \\
&&+(\alpha _{2}(x_{2})-\tilde{v}_{1}(x))K_{+,11}.
\end{eqnarray*}%
Moreover, the following conditions
\begin{eqnarray}
\tilde{u}_{1}(x) &=&-K_{+,12}^{(0)}|_{y=x},\text{ \ \ }\tilde{u}%
_{2}(x)=-K_{+,21}^{(0)}|_{y=x},\text{ }  \label{3.22} \\
\tilde{v}_{2}(x)|_{x_{1}=-\infty } &=&\alpha _{2}(x_{2}),\text{ \ }\tilde{v}%
_{1}(x)|_{x_{2}=-\infty }=\alpha _{1}(x_{1})\text{ }  \notag
\end{eqnarray}%
hold for all $x\in \mathbb{R}^{2}$ and \ $y\in $supp$\hat{K}_{+}(\mathbf{%
\Omega }),$ where we take into account the singular series expansion%
\begin{equation}
\hat{K}_{+}(\mathbf{\Omega })=\sum_{s=0}^{p(K_{+})}K_{+}^{(s)}\delta
_{\sigma _{x}^{(1)}}^{(s-1)}  \label{3.22a}
\end{equation}%
for some finite integer $p(K_{+})\in \mathbb{Z}_{+}$ with respect to the
Dirac function $\delta _{\sigma _{x}^{(1)}}:W_{2}^{q}(\mathbb{R}^{2};\mathbb{%
C})$\bigskip $\rightarrow \mathbb{R},$ $q\in \mathbb{Z}_{+},$ and its
derivatives, having the support (see \cite{GS}, Chapter 3) coinciding with
the closed cycle $\sigma _{x}^{(1)}\in \mathcal{K}(\mathbb{R}^{2}).$

\begin{remark}
Concerning the special case (\ref{3.13}) discussed before in \cite{Ni,NP},
one gets easily that $p(K_{+})=1$ and $\sigma _{x}^{(1)}=\partial (\cap _{j=%
\overline{1,2}}\{y\in \mathbb{R}^{2}:<y-x,\gamma _{j}>=0\})\subset $supp$%
\hat{K}_{+}(\mathbf{\Omega }).$ It was shown also before that equations like
(\ref{3.21}) and (\ref{3.22}) possess solutions if the
Gelfand-Levitan-Marchenko equation (\ref{eq:2.25}) \ does.
\end{remark}

Making use also of the exact forms of operators \textrm{L}$_{1}$ and \textrm{%
L}$_{2}\in \mathcal{L(H)},$ one obtains easily from (\ref{3.14}) and (\ref%
{3.19}) the corresponding set of differential equations for components of
the kernel $\hat{\Phi}(\mathbf{\Omega })\in H_{-}\otimes H_{-}:$%
\begin{eqnarray}
\frac{\partial \Phi _{11}}{\partial x_{1}}+\frac{\partial \Phi _{11}}{%
\partial y_{1}} &=&0,\text{ \ }\frac{\partial \Phi _{12}}{\partial x_{1}}+%
\frac{\partial \Phi _{12}}{\partial y_{1}}=0,  \label{3.23} \\
\frac{\partial \Phi _{21}}{\partial x_{2}}+\frac{\partial \Phi _{21}}{%
\partial y_{2}} &=&0,\text{ \ }\frac{\partial \Phi _{22}}{\partial x_{2}}+%
\frac{\partial \Phi _{22}}{\partial y_{2}}=0,  \notag
\end{eqnarray}%
\begin{eqnarray*}
\frac{\partial \Phi _{11}}{\partial t}\pm (\frac{\partial ^{2}}{\partial
x_{2}^{2}}-\frac{\partial ^{2}}{\partial y_{2}^{2}})\Phi _{11}+(\alpha
_{2}(y_{2})-\alpha _{2}(x_{2}))\Phi _{11} &=&0, \\
\frac{\partial \Phi _{12}}{\partial t}\pm (\frac{\partial ^{2}}{\partial
x_{2}^{2}}-\frac{\partial ^{2}}{\partial y_{2}^{2}})\Phi _{12}+(\alpha
_{1}(y_{1})-\alpha _{2}(x_{2}))\Phi _{12} &=&0, \\
\frac{\partial \Phi _{21}}{\partial t}+(\frac{\partial ^{2}}{\partial
x_{1}^{2}}-\frac{\partial ^{2}}{\partial y_{1}^{2}})\Phi _{21}+(\alpha
_{2}(y_{2})-\alpha _{1}(x_{1}))\Phi _{21} &=&0, \\
\frac{\partial \Phi _{22}}{\partial t}+(\frac{\partial ^{2}}{\partial
x_{1}^{2}}-\frac{\partial ^{2}}{\partial y_{1}^{2}})\Phi _{22}+(\alpha
_{1}(y_{1})-\alpha _{1}(x_{1}))\Phi _{22} &=&0
\end{eqnarray*}%
for all (x,y)$\in \mathbb{R}^{2}\times \mathbb{R}^{2}.$The obtained above
equations (\ref{3.23}) generalize those before found in \cite{Ni,NP} and
used for exact integrating the well known Devey-Stewartson differential
equation \cite{ZS,No,FT} and finding so called soliton like solutions.
Concerning our generalized case the kernel (\ref{3.22a}) is a solution to
the following Gelfand-Levitan-Marchenko type equations:%
\begin{equation*}
K_{+}^{(0)}(x;y)+\Phi ^{(0)}(x;y)+\int_{S_{+}^{(2)}(\sigma _{x}^{(1)},\sigma
_{\infty }^{(1)})}K_{+}^{(0)}(x;\xi )\Phi ^{(0)}(\xi ;y)d\xi
\end{equation*}%
\begin{equation}
+\int_{\sigma _{x}^{(1)}}K_{+}^{(1)}(x;\xi )\Phi ^{(0)}(\xi ;y)d\sigma
_{x}^{(1)}=0,\text{ \ }  \label{3.24}
\end{equation}%
\begin{equation*}
K_{+}^{(1)}(x;y)+\Phi ^{(1)}(x;y)+\int_{S_{+}^{(2)}(\sigma _{x}^{(1)},\sigma
_{\infty }^{(1)})}K_{+}^{(0)}(x;\xi )\Phi ^{(1)}(\xi ;y)d\xi
\end{equation*}%
\begin{equation*}
+\int_{\sigma _{x}^{(1)}}K_{+}^{(1)}(x;\xi )\Phi ^{(1)}(\xi ;y)d\sigma
_{x}^{(1)}=0,
\end{equation*}%
where $y\in S_{+}^{(2)}(\sigma _{x}^{(1)},\sigma _{\infty }^{(1)})$ for all $%
x\in \mathbb{R}^{2}$ and, by definition,
\begin{equation}
\hat{\Phi}(\mathbf{\Omega }):=\Phi ^{(0)}+\Phi ^{(1)}\delta _{\sigma
_{x}^{(1)}}  \label{3.25}
\end{equation}%
is the corresponding to (\ref{3.22a}) kernel expansion. Since the kernel (%
\ref{3.25}) is singular, the differential equations (\ref{3.23}) must be
treated naturally in the distributional sense \cite{GS}.

Taking into account the exact forms of "dressed" differential operators $%
\mathrm{L}_{j}\in \mathcal{L(H)},$ $j=\overline{1,2},$ given by (\ref{3.10})
and (\ref{3.15}) one gets easily that the commutativity condition (\ref{3.18}%
) gives rise to that of $\mathrm{\tilde{L}}_{j}\in \mathcal{L(H)},$ $j=%
\overline{1,2},$ being equivalent to the mentioned before
Devey-Stewartson\bigskip\ dynamical system%
\begin{eqnarray}
d\tilde{u}_{1}/dt &=&-(\tilde{u}_{1,xx}+\tilde{u}_{1,yy})+2(\tilde{v}_{1}-%
\tilde{v}_{2}),  \label{3.26} \\
d\tilde{u}_{2}/dt &=&\tilde{u}_{2,xx}+\tilde{u}_{2,yy}+2(\tilde{v}_{2}-%
\tilde{v}_{1}),  \notag \\
\tilde{v}_{1,x} &=&(\tilde{u}_{1}\tilde{u}_{2})_{y},\text{ \ }\tilde{v}%
_{2,x}=(\tilde{u}_{1}\tilde{u}_{2})_{x}\text{ }  \notag
\end{eqnarray}%
on a functional infinite-dymensioanl manifold $M_{u}\subset \mathcal{S}(%
\mathbb{R}^{2};\mathbb{C}).$ The exact soliton like solutions to (\ref{3.26}%
) are given by expressions (\ref{3.22}), where the kernel $K_{+}^{(1)}(%
\mathbf{\Omega })$ solves the second linear integral equation of (\ref{3.24}%
). On the other hand-side, there exists the exact expression (\ref{eq:2.4})
which solves the set of "dressed" equations
\begin{equation}
\mathrm{\tilde{L}}_{1}\tilde{\psi}^{(0)}(\eta )=0,\text{ \ }\mathrm{\tilde{L}%
}_{2}\tilde{\psi}^{(0)}(\eta )=0.  \label{3.27}
\end{equation}%
Since the kernels $\Omega (\lambda ,\mu )\in L_{2}^{(\rho )}(\Sigma ;\mathbb{%
C})\otimes L_{2}^{(\rho )}(\Sigma ;\mathbb{C}),$ for $\lambda ,\mu \in
\Sigma ,$ $(t;x)\in M_{\mathrm{T}}\cap $ $S_{+}^{(2)}(\sigma
_{x}^{(1)},\sigma _{\infty }^{(1)})$ are given by means of exact expressions
(\ref{eq:2.2}), one can find via simple calculations the corresponding
analytical expression for the functions $(\tilde{u}_{1},\tilde{u}_{2})\in
M_{u},$ solving the dynamical system (\ref{3.26}). This procedure is often
called the Darboux type transformation and was recently extensively used as
a particular case of the construction above in \cite{SP} for finding
soliton-like solutions to the Devey-Stewartson (\ref{3.26}) and related with
it two-dimensional modified Korteweg-de Vries flows on $M_{u.}$ Moreover,
\bigskip\ as it can be observed from the technique used for constructing the
Delsarte-Darboux \bigskip transmutation operators $\mathbf{\Omega }_{\pm
}\in \mathcal{B(H)},$ the set of solutions to (\ref{3.26}) \bigskip obtained
by means of Darboux type transformations coincides completely with the
corresponding set of solutions obtained by means of solving the related set
of Gelfand-Levitan-Marchenko integral equations (\ref{3.23}) and (\ref{3.24}%
).

\textbf{2.3 An affine de Rham-Hodge-Skrypnik differential complex and
related generalized self-dual Yang-Mills flows.} Consider the following set
of affine differential expressions in $\mathcal{H}:=C^{1}(\mathbb{R}%
^{m+1};H),$ $H:=L_{2}(\mathbb{R}^{m};\mathbb{C}^{N}):$%
\begin{equation}
\mathrm{L}_{i}(\lambda ):=\mathbf{1}\frac{\partial }{\partial p_{i}}-\lambda
\frac{\partial }{\partial x_{i}}+A_{i}(x;p|t),  \label{3.28}
\end{equation}%
where $x\in \mathbb{R}^{m},(t,p)\in \mathbb{R}^{m+1},$ matrices $A_{i}\in
C^{1}(\mathbb{R}^{m+1};S(\mathbb{R}^{m};End\mathbb{C}^{N})),$ $i=\overline{%
1,m},$ and a parameter $\lambda \in \mathbb{C}.$ One can$\mathbb{\ }$easily
now construct an exact affine de Rham-Hodge-Skrypnik differential complex on
$M_{\mathrm{T}}:=\mathbb{R}^{m+1}\mathbb{\times R}^{m}$ as%
\begin{equation}
\mathcal{H}\rightarrow \Lambda (M_{\mathrm{T}};\mathcal{H)}\overset{d_{%
\mathcal{L}(\lambda )}}{\rightarrow }\Lambda ^{1}(M_{\mathrm{T}};\mathcal{H}%
)\rightarrow \overset{d_{\mathcal{L}(\lambda )}}{...}\rightarrow \Lambda
^{2m+1}(M_{\mathrm{T}};\mathcal{H})\overset{d_{\mathcal{L}(\lambda )}}{%
\rightarrow }0,  \label{3.29}
\end{equation}%
where, by definition, the differentiation
\begin{equation}
d_{\mathcal{L}(\lambda )}:=dt\wedge \mathrm{B}(\lambda
)+\sum_{i=1}^{m}dp_{i}\wedge \mathrm{L}_{i}(\lambda )  \label{3.30}
\end{equation}%
and the affine matrix%
\begin{equation}
\mathrm{B}(\lambda ):=\partial /\partial
t-\sum_{s=0}^{n(B)+q}B_{s}(x;p|t)\lambda ^{n(B)-s}  \label{3.31}
\end{equation}%
with matrices $B_{s}\in C^{1}(\mathbb{R}^{m+1};S(\mathbb{R}^{m};End\mathbb{C}%
^{N})),$ $s=\overline{0,n(B)+q},$ $n(B),q\in \mathbb{Z}_{+}.$ The affine
complex (\ref{3.29}) will be exact for all $\lambda \in \mathbb{C}$ iff the
following generalized self-dual Yang-Mills equations \cite{Gu}
\begin{equation*}
\partial A_{i}/\partial p_{j}-\partial A_{j}/\partial p_{i}-[A_{i},A_{j}]=0,%
\text{ \ }\partial A_{i}/\partial x_{j}-\partial A_{j}/\partial x_{i}=0,
\end{equation*}%
\begin{equation*}
\partial B_{0}/\partial x_{i}=0,\text{ }\partial B_{n(B)+q}/\partial p_{i}=0,%
\text{ }\partial B_{s}/\partial x_{i}=\partial B_{s-1}/\partial
p_{i}+[A_{i},B_{s-1}]=0,
\end{equation*}%
\begin{equation}
\partial A_{i}/\partial t+\partial B_{n(B)}/\partial p_{i}-\partial
B_{n(B)+1}/\partial x_{i}+[A_{i},B_{n(B)}]=0  \label{3.32}
\end{equation}%
hold for all $i,j=\overline{1,m}$ and $s=\overline{0,n(B)}\vee \overline{%
n(B)+q,n(B)+2}.$ Assume now that the conditions (\ref{3.32}) are satisfied
on $M_{\mathrm{T}}.$ Then, making the change $\mathbb{C\ni }\lambda
\rightarrow \partial /\partial \tau :\mathcal{H}\rightarrow \mathcal{H},$ $%
\tau \in \mathbb{R},$ one finds the following set of pure differential
expressions%
\begin{eqnarray}
\mathrm{L}_{i(\tau )} &:&=\mathbf{1}\frac{\partial }{\partial p_{i}}-\frac{%
\partial ^{2}}{\partial \tau \partial x_{i}}+A_{i}(x;p|t),  \label{3.33} \\
\mathrm{B}_{(\tau )} &:&=\partial /\partial
t-\sum_{s=0}^{n(B)+q}B_{s}(x;p|t)(\frac{\partial }{\partial \tau })^{n(B)-s},
\notag
\end{eqnarray}%
where matrices $A_{i},$ $i=\overline{1,m},$ and $B_{s},$ $s=\overline{%
0,n(B)+q},$ don't depend on the variable $\tau \in \mathbb{R}.$ By means
operator expressions (\ref{3.33}) one can now naturally construct a new
differential complex related with that of (\ref{3.29}):%
\begin{equation}
\mathcal{H}_{(\tau )}\rightarrow \Lambda (M_{\mathrm{T,\tau }};\mathcal{H}%
_{(\tau )}\mathcal{)}\overset{d_{\mathcal{L}}}{\rightarrow }\Lambda ^{1}(M_{%
\mathrm{T,\tau }};\mathcal{H}_{(\tau )})\rightarrow \overset{d_{\mathcal{L}}}%
{...}\rightarrow \Lambda ^{2m+2}(M_{\mathrm{T,\tau }};\mathcal{H}_{(\tau )})%
\overset{d_{\mathcal{L}}}{\rightarrow }0,  \label{3.34}
\end{equation}%
where, by definition, $\mathcal{H}_{(\tau )}:=C^{1}(\mathbb{R}%
^{m+1};H_{(\tau )}),$ $H_{(\tau )}:=L_{2}(\mathbb{R}^{m}\times \mathbb{R}%
_{\tau };\mathbb{C}^{N})$ and

\bigskip
\begin{equation}
d_{\mathcal{L}}:=dt\wedge \mathrm{B}_{(\tau )}+\sum_{i=1}^{m}dp_{i}\wedge
\mathrm{L}_{i(\tau )}.  \label{3.35}
\end{equation}%
Owing to the condition (\ref{3.32}) the following lemma holds.

\begin{lemma}
The differential complex (\ref{3.34}) is exact.
\end{lemma}

Therefore, one can build the standard de Rham-Hodge-Skrypnik type Hilbert
space decomposition%
\begin{equation}
\mathcal{H}_{\Lambda }(M_{\mathrm{T,\tau }}):=\overset{k=2m+2}{\underset{k-0}%
{\oplus }}\mathcal{H}_{\Lambda }^{k}(M_{\mathrm{T,\tau }})  \label{3.36}
\end{equation}%
as well the corresponding Hilbert-Schmidt rigging
\begin{equation}
\mathcal{H}_{\Lambda ,+}(M_{\mathrm{T,\tau }})\subset \mathcal{H}_{\Lambda
}(M_{\mathrm{T,\tau }})\subset \mathcal{H}_{\Lambda ,-}(M_{\mathrm{T,\tau }%
}).  \label{3.37}
\end{equation}%
Making use now of the results obtained in subsection 1.5, one can define the
Delsarte closed subspaces $\mathcal{H}_{0(\tau )}$ and $\mathcal{\tilde{H}}%
_{0(\tau )}\subset \mathcal{H}_{(\tau )-},$ related with the exact complex (%
\ref{3.34}):%
\begin{eqnarray}
\mathcal{H}_{0(\tau )} &:&=\{\psi _{(\tau )}^{(0)}(\xi )\in \mathcal{H}%
_{\Lambda ,-}^{0}(M_{\mathrm{T,\tau }}):\mathrm{L}_{j(\tau )}\psi _{(\tau
)}^{(0)}(\xi )=0,  \label{3.38} \\
\text{ }\mathrm{B}_{(\tau )}\psi _{(\tau )}^{(0)}(\xi ) &=&0,\text{ }\psi
_{(\tau )}^{(0)}(\xi )|_{\Gamma }=0,\text{ }\psi _{(\tau )}^{(0)}(\xi
)|_{t=0}=e^{\lambda \tau }\psi _{\lambda }^{(0)}(\eta )\in \mathcal{H}%
_{\Lambda ,-}^{0}(M_{\mathbb{R}^{m}\mathrm{,\tau }}),  \notag \\
\text{ }\mathrm{L}_{j}(\lambda )\psi _{\lambda }^{(0)}(\eta ) &=&0,\text{ }%
\xi =(\lambda ;\eta )\in \Sigma :\mathbb{=C}\times \Sigma _{\mathbb{C}%
}^{(m)}\},  \notag
\end{eqnarray}%
\bigskip
\begin{eqnarray*}
\mathcal{\tilde{H}}_{0(\tau )} &:&=\{\tilde{\psi}_{(\tau )}^{(0)}(\xi )\in
\mathcal{H}_{\Lambda ,-}^{0}(M_{\mathrm{T,\tau }}):\mathrm{\tilde{L}}%
_{j(\tau )}^{(0)}\tilde{\psi}_{(\tau )}^{(0)}(\xi )=0, \\
\text{ }\mathrm{\tilde{B}}_{(\tau )}\tilde{\psi}_{(\tau )}^{(0)}(\xi ) &=&0,%
\text{ }\tilde{\psi}_{(\tau )}^{(0)}(\xi )|_{\tilde{\Gamma}}=0,\text{ }%
\tilde{\psi}_{(\tau )}^{(0)}(\xi )|_{t=0}=e^{\lambda \tau }\tilde{\psi}%
_{\lambda }^{(0)}(\eta )\in \mathcal{H}_{\Lambda ,-}^{0}(M_{\mathbb{R}^{m}%
\mathrm{,\tau }}), \\
\text{ }\mathrm{\tilde{L}}_{j}(\lambda )\tilde{\psi}_{\lambda }^{(0)}(\eta )
&=&0,\text{ }\xi =(\lambda ;\eta )\in \Sigma :\mathbb{=C}\times \Sigma _{%
\mathbb{C}}^{(m)}\},
\end{eqnarray*}%
where $\Gamma $ and $\tilde{\Gamma}\subset M_{\mathrm{T,\tau }}$\bigskip\
are some smooth hyper-surfaces. The similar expressions correspond to the
adjoint closed subspaces $\mathcal{H}_{0(\tau )}^{\ast }$ and $\mathcal{%
\tilde{H}}_{0(\tau )}^{\ast }\subset \mathcal{H}_{\tau ,-}^{\ast }:$\bigskip
\begin{eqnarray}
\mathcal{\tilde{H}}_{0(\tau )} &:&=\{\varphi _{(\tau )}^{(0)}(\xi )\in
\mathcal{H}_{\Lambda ,-}^{0}(M_{\mathrm{T,\tau }}):\mathrm{L}_{j(\tau
)}^{\ast }\varphi _{(\tau )}^{(0)}(\xi )=0,  \label{3.39} \\
\text{ }\mathrm{B}_{(\tau )}\varphi _{(\tau )}^{(0)}(\xi ) &=&0,\text{ }%
\varphi _{(\tau )}^{(0)}(\xi )|_{\Gamma }=0,\text{ }\varphi _{(\tau
)}^{(0)}(\xi )|_{t=0}=e^{-\bar{\lambda}\tau }\varphi _{\lambda }^{(0)}(\eta
)\in \mathcal{H}_{\Lambda ,-}^{0}(M_{\mathbb{R}^{m}\mathrm{,\tau }}),  \notag
\\
\text{ }\mathrm{L}_{j}^{\ast }(\lambda )\varphi _{\lambda }^{(0)}(\eta )
&=&0,\text{ }\xi =(\lambda ;\eta )\in \Sigma :\mathbb{=C}\times \Sigma _{%
\mathbb{C}}^{(m)}\},  \notag
\end{eqnarray}%
\bigskip
\begin{eqnarray*}
\mathcal{\tilde{H}}_{0(\tau )} &:&=\{\tilde{\varphi}_{(\tau )}^{(0)}(\xi
)\in \mathcal{H}_{\Lambda ,-}^{0}(M_{\mathrm{T,\tau }}):\mathrm{\tilde{L}}%
_{j(\tau )}^{\ast }\tilde{\varphi}_{(\tau )}^{(0)}(\xi )=0, \\
\text{ }\mathrm{\tilde{B}}_{(\tau )}^{\ast }\tilde{\varphi}_{(\tau
)}^{(0)}(\xi ) &=&0,\text{ }\tilde{\varphi}_{(\tau )}^{(0)}(\xi )|_{\tilde{%
\Gamma}}=0,\text{ }\tilde{\varphi}_{(\tau )}^{(0)}(\xi )|_{t=0}=e^{-\bar{%
\lambda}\tau }\tilde{\varphi}_{\lambda }^{(0)}(\eta )\in \mathcal{H}%
_{\Lambda ,-}^{0}(M_{\mathbb{R}^{m}\mathrm{,\tau }}), \\
\text{ }\mathrm{\tilde{L}}_{j}^{\ast }(\lambda )\tilde{\varphi}_{\lambda
}^{(0)}(\eta ) &=&0,\text{ }\xi =(\lambda ;\eta )\in \Sigma :\mathbb{=C}%
\times \Sigma _{\mathbb{C}}^{(m)}\}.
\end{eqnarray*}%
Based on the closed subspaces (\ref{3.39}) and (\ref{3.38}), one can
suitably build the Darboux type kernel $\tilde{\Omega}_{(t,x;\tau )}(\eta
,\xi )\in L_{2}^{(\rho )}(\Sigma _{\mathbb{C}}^{(m)};\mathbb{C})\otimes
L_{2}^{(\rho )}(\Sigma _{\mathbb{C}}^{(m)};\mathbb{C}),$ $\eta ,\xi \in
\Sigma _{\mathbb{C}}^{(m)},$ and further, the corresponding Delsarte
transmutation mappings $\mathbf{\Omega }_{\pm }\in \mathcal{B}(H_{(\tau )}).$
Namely, assume that the following conditions%
\begin{equation}
\psi _{(\tau )}^{(0)}(\xi ):=\tilde{\psi}_{(\tau )}^{(0)}(\xi )\cdot \tilde{%
\Omega}_{(t,p;x;\tau )}^{-1}\tilde{\Omega}_{(t_{0},p_{0,}x_{0};\tau )}
\label{3.40}
\end{equation}%
for any $\xi \in \mathbb{C\times }\Sigma _{\mathbb{C}}^{(m)}$ hold, where%
\begin{equation*}
\tilde{\Omega}_{(t,x;\tau )}(\mu ,\xi ):=\int_{\sigma _{(t;x;\tau )}}\tilde{%
\Omega}_{(\tau )}^{(2m+1)}[e^{-\bar{\lambda}\tau }\tilde{\varphi}^{(0)}(\mu
),e^{\lambda \tau }\tilde{\psi}^{(0)}(\eta )dx\wedge dp\wedge dt],
\end{equation*}%
\begin{eqnarray}
&&\tilde{Z}_{(\tau )}^{(2m+1)}[e^{-\bar{\lambda}\tau }\tilde{\varphi}%
^{(0)}(\mu ),\sum_{i=1}^{m}e^{\lambda \tau }\tilde{\psi}^{(0)}(\xi
_{(i)})\wedge d\tau \wedge dx\underset{j\neq i}{\overset{m}{\wedge }}dp_{j}]
\label{3.41} \\
&:&=d\tilde{\Omega}_{(\tau )}^{(2m)}[e^{-\bar{\lambda}\tau }\tilde{\varphi}%
^{(0)}(\mu ),\sum_{i=1}^{m}e^{\lambda \tau }\tilde{\psi}^{(0)}(\xi
_{(i)})\wedge d\tau \wedge dx\underset{j\neq i}{\overset{m}{\wedge }}dp_{j}],
\notag
\end{eqnarray}%
and, similarly to (\ref{eq:1.24}), there holds the relationship \ \
\begin{eqnarray}
&<&d_{\mathcal{\tilde{L}}}^{\ast }\tilde{\varphi}^{(0)}(\mu )e^{-\bar{\lambda%
}\tau },\ast \sum_{i=1}^{m}e^{\lambda \tau }\tilde{\psi}^{(0)}(\xi
_{(i)})dt\wedge d\tau \wedge dx\underset{j\neq i}{\overset{m}{\wedge }}%
dp_{j}>  \label{3.42} \\
&=&<(\ast )^{-1}\tilde{\varphi}^{(0)}(\mu )e^{-\bar{\lambda}\tau },d_{%
\mathcal{\tilde{L}}}(\sum_{i=1}^{m}e^{\lambda \tau }\tilde{\psi}^{(0)}(\xi
_{(i)})dt\wedge d\tau \wedge dx\underset{j\neq i}{\overset{m}{\wedge }}%
dp_{j})>  \notag \\
&&+d\tilde{Z}_{(\tau )}^{(2m+1)}[\tilde{\varphi}^{(0)}(\mu )e^{-\bar{\lambda}%
\tau },\sum_{i=1}^{m}e^{\lambda \tau }\tilde{\psi}^{(0)}(\xi _{(i)})dt\wedge
d\tau \wedge dx\underset{j\neq i}{\overset{m}{\wedge }}dp_{j}],  \notag
\end{eqnarray}%
defining the exact ($2m+1)$-form $\tilde{Z}_{(\tau )}^{(2m+1)}\in \Lambda
^{2m+1}(M_{\mathrm{T},\tau };\mathbb{C}).$ Compute now the Delsarte
transformed differential expressions
\begin{equation}
\mathrm{L}_{j(\tau )}:=\mathbf{\hat{\Omega}}_{(\tau )\pm }^{-1}\mathrm{%
\tilde{L}}_{j(\tau )}\mathbf{\hat{\Omega}}_{(\tau )\pm },\text{ \ }\mathrm{B}%
_{(\tau )}:=\mathbf{\hat{\Omega}}_{(\tau )\pm }^{-1}\mathrm{\tilde{B}}%
_{(\tau )}\mathbf{\hat{\Omega}}_{(\tau )\pm }\text{\ }  \label{3.43}
\end{equation}%
for any $j=\overline{1,m},$ where, by definition, \bigskip
\begin{eqnarray}
\mathrm{\tilde{L}}_{j(\tau )} &:&=\mathbf{1}\frac{\partial }{\partial p_{j}}-%
\frac{\partial ^{2}}{\partial \tau \partial x_{j}}+\bar{A}_{j},  \label{3.44}
\\
\mathrm{B}_{(\tau )} &:&=\partial /\partial t-\sum_{s=0}^{n(B)+q}\bar{B}_{s}(%
\frac{\partial }{\partial \tau })^{n(B)-s}  \notag
\end{eqnarray}%
with all matrices $\bar{A}_{j}\in End\mathbb{C}^{m},$ $j=\overline{1,m},$
and $\bar{B}_{s}\in End\mathbb{C}^{m},$ $s=\overline{0,n(B)+q},$ being
constant. This means, in particular, the commuting relationships
\begin{equation}
\lbrack \mathrm{\tilde{L}}_{j(\tau )},\mathrm{\tilde{L}}_{i(\tau )}]=0,\text{
}[\mathrm{\tilde{L}}_{j(\tau )},\mathrm{\tilde{B}}_{(\tau )}]=0  \label{3.45}
\end{equation}%
hold for all $i,j=\overline{1,m}.$ Owing to the expressions (\ref{3.43}) the
induced commuting relationships
\begin{equation}
\lbrack \mathrm{L}_{j(\tau )},\mathrm{L}_{i(\tau )}]=0,\text{ }[\mathrm{L}%
_{j(\tau )},\mathrm{B}_{(\tau )}]=0  \label{3.46}
\end{equation}%
evidently hold, coinciding exactly with relationships (\ref{3.32}).
Moreover, reducing our differential expressions (\ref{3.43}) upon functional
subspaces $\mathcal{H}_{(\lambda )}:=e^{\lambda \tau }\mathcal{H},$ $\lambda
\in \mathbb{C},$ one gets easily the set of affine differential expressions (%
\ref{3.28}) and (\ref{3.31}). Write down now the respectively reduced
Delsarte transmutation operators
\begin{eqnarray}
\mathbf{\hat{\Omega}}_{\pm } &=&\mathbf{1-}\int_{\Sigma _{\mathbb{C}%
}^{(m)}}d\rho _{\Sigma _{\mathbb{C}}^{(m)}}(\nu )\int_{\Sigma _{\mathbb{C}%
}^{(m)}}d\rho _{\Sigma _{\mathbb{C}}^{(m)}}(\eta )\psi ^{(0)}(\lambda ;\nu )%
\tilde{\Omega}_{(t_{0},p_{0};x_{0})}^{-1}(\lambda ;\nu ,\eta )  \notag \\
&&\times \int_{S_{\pm }^{(2m+1)}(\sigma _{(t,p;x)}^{(2m)},\sigma
_{(tt_{0},p_{0};x_{0})}^{(2m)})}\tilde{Z}^{(2m+1)}[\tilde{\varphi}%
^{(0)}(\lambda ;\nu ),(\cdot )\sum_{i=1}^{m}dt\wedge dx\underset{j\neq i}{%
\overset{m}{\wedge }}dp_{j}],  \label{3.47}
\end{eqnarray}%
where $\sigma _{(t,p;x)}^{(2m)}$ and $\sigma
_{(tt_{0},p_{0};x_{0})}^{(2m)}\in \mathcal{K}(M_{\mathrm{T}})$ are some $2m$%
-dimensional closed singular simplexes, and by definition,%
\begin{eqnarray*}
&&\tilde{Z}^{(2m+1)}[\tilde{\varphi}^{(0)}(\lambda ;\nu ),\sum_{i=1}^{m}%
\tilde{\psi}^{(0)}(\lambda ;\eta _{(i)})dt\wedge dx\underset{j\neq i}{%
\overset{m}{\wedge }}dp_{j}] \\
&:&=\tilde{Z}_{(\tau )}^{(2m+1)}[e^{-\bar{\lambda}\tau }\tilde{\varphi}%
^{(0)}(\lambda ;\nu ),\sum_{i=1}^{m}e^{\lambda \tau }\tilde{\psi}%
^{(0)}(\lambda ;\eta _{(i)})d\tau \wedge dt\wedge dx\underset{j\neq i}{%
\overset{m}{\wedge }}dp_{j}]|_{d\tau =0},
\end{eqnarray*}%
\begin{equation}
d\tilde{\Omega}_{(t,p;x)}(\lambda ;\nu ,\eta ):=\tilde{Z}^{(2m+1)}[\tilde{%
\varphi}^{(0)}(\lambda ;\nu ),\sum_{i=1}^{m}\tilde{\psi}^{(0)}(\lambda ;\eta
_{(i)})dt\wedge dx\underset{j\neq i}{\overset{m}{\wedge }}dp_{j}],
\label{3.48}
\end{equation}%
since the $(2m+1)$-form (\ref{3.48}) is owing to (\ref{3.42}) also exact for
any $(\lambda ;\nu ,\eta )\in \mathbb{C\times (}\mathbb{\Sigma }_{%
\mathbb{C}}^{(m)}\times \mathbb{\Sigma }_{\mathbb{C}}^{(m)}).$
Thus, the operator expression (\ref{3.47}) if applied to the
operators (\ref{3.44}) reduced upon the functional subspace
$\mathcal{H}_{(\lambda )}\simeq \mathcal{H},$ $\lambda \in
\mathbb{C},$ gives rise to the differential expressions
\begin{equation}
\mathrm{L}_{j}(\lambda ):=\mathbf{\hat{\Omega}}_{\pm }^{-1}\mathrm{\tilde{L}}%
_{j}(\lambda )\mathbf{\hat{\Omega}}_{\pm }\text{ \ }\mathrm{B}(\lambda ):=%
\mathbf{\hat{\Omega}}_{\pm }^{-1}\mathrm{\tilde{B}}(\lambda )\mathbf{\hat{%
\Omega}}_{\pm },  \label{3.49}
\end{equation}%
where $\mathrm{L}_{j}(\lambda )\mathcal{H}_{(\lambda )}=\mathrm{L}_{j(\tau )}%
\mathcal{H}_{(\lambda )},$ $\mathrm{B}(\lambda )\mathcal{H}_{(\lambda )}=%
\mathrm{B}_{(\tau )}(\lambda )\mathcal{H}_{(\lambda )},$ $j=\overline{1,m},$
coinciding with affine differential expressions (\ref{3.28}) and (\ref{3.31}%
). Concerning application of these results to finding exact soliton like
solutions to self-dual Yang-Mills equations (\ref{3.32}), it is enough to
mention that the relationship (\ref{3.40}) reduced upon the subspace $%
\mathcal{H}_{(\lambda )}\simeq \mathcal{H},$ $\lambda \in \mathbb{C},$ gives
rise the following mapping:
\begin{equation}
\psi ^{(0)}(\lambda ;\eta ):=\tilde{\psi}^{(0)}(\lambda ;\eta )\cdot \tilde{%
\Omega}_{(t,p;x)}^{-1}\tilde{\Omega}_{(t_{0},p_{0};x_{0})},  \label{3.50}
\end{equation}%
where kernels $\tilde{\Omega}_{(t,p;x;\tau )}(\lambda ;\eta ,\xi )\in
L_{2}^{(\rho )}(\Sigma _{\mathbb{C}}^{(m)};\mathbb{C})\otimes L_{2}^{(\rho
)}(\Sigma _{\mathbb{C}}^{(m)};\mathbb{C}),$ $\eta ,\xi \in \Sigma _{\mathbb{C%
}}^{(m)},$ for all $(t,p;x)\in M_{\mathrm{T}}$ and $\lambda \in \mathbb{C}.$
Since the element $\psi ^{(0)}(\lambda ;\eta )\in \mathcal{H}_{-}$ for any $%
(\lambda ;\xi )\in \mathbb{C\times }\Sigma _{\mathbb{C}}^{(m)}$ satisfies
the set of differential equations%
\begin{equation}
{\normalsize L}_{i}(\lambda )\psi ^{(0)}(\lambda ;\eta )=0,\text{ \ }%
{\normalsize B}(\lambda )\psi ^{(0)}(\lambda ;\eta )=0,  \label{3.51}
\end{equation}%
for all $i=\overline{1,m},$ from (\ref{3.50}) and (\ref{3.51}) one finds
easily exact expressions for the corresponding matrices $A_{j}$ and $%
B_{s}\in C^{1}(\mathbb{R}\times \mathbb{R}^{m+1};S(\mathbb{R}^{m};End\mathbb{%
C}^{N})),$ $j=\overline{1,m},$ $s=\overline{0,n(B)+q},$ satisfying the
self-dual Yang-Mills equations (\ref{3.32}). Thereby, the following theorem
is stated.

\begin{theorem}
The integral expressions (\ref{3.47}) in $\mathcal{H}$ are the Delsarte
transmutation operators corresponding to the affine differential expressions
(\ref{3.28}), (\ref{3.32}) and constant operators
\begin{equation}
\mathrm{\tilde{L}}_{i}(\lambda ):=\mathbf{1}\frac{\partial }{\partial p_{i}}%
-\lambda \frac{\partial }{\partial x_{i}}+\bar{A},\text{ \ }\mathrm{\tilde{B}%
}(\lambda ):=\partial /\partial t-\sum_{s=0}^{n(B)+q}\bar{B}_{s}\lambda
^{n(B)-s}  \label{3.52}
\end{equation}%
for any $\lambda \in \mathbb{C}.$ The mapping (\ref{3.50}) realizes the
isomorphisms between the closed subspaces%
\begin{eqnarray}
\mathcal{H}_{0} &:&=\{\psi ^{(0)}(\lambda ;\eta )\in \mathcal{H}_{-}:d_{%
\mathcal{\tilde{L}(\lambda )}}\psi ^{(0)}(\lambda ;\eta )=0,\text{ }\psi
^{(0)}(\lambda ;\eta )|_{t=0}  \label{3.53} \\
&=&\psi _{\lambda }^{(0)}(\eta )\in H_{-},\text{ }\psi ^{(0)}(\lambda ;\eta
)|_{\Gamma }=0,(\lambda ;\eta )\in \mathbb{C}\times \Sigma _{\mathbb{C}%
}^{(m)}\}  \notag
\end{eqnarray}%
and
\begin{eqnarray}
\mathcal{\tilde{H}}_{0} &:&=\{\tilde{\psi}^{(0)}(\lambda ;\eta )\in \mathcal{%
H}_{-}:d_{\mathcal{\tilde{L}(\lambda )}}^{(0)}\tilde{\psi}(\lambda ;\eta )=0,%
\text{ }\tilde{\psi}^{(0)}(\lambda ;\eta )|_{t=0}  \label{3.54} \\
&=&\tilde{\psi}_{\lambda }^{(0)}(\eta )\in H_{-},\text{ }\tilde{\psi}%
^{(0)}(\lambda ;\eta )|_{\tilde{\Gamma}}=0,(\lambda ;\eta )\in \mathbb{%
C\times }\Sigma _{\mathbb{C}}^{(m)}\}  \notag
\end{eqnarray}%
for any parameter $\lambda \in \mathbb{C}.$ Moreover, the expressions (\ref%
{3.50}) generate the standard Darboux type transformations for the set of
operators (\ref{3.52}) and (\ref{3.28}), (\ref{3.31}) via the corresponding
set of linear equations (\ref{3.51}), thereby producing exact soliton-like
solutions to the self-dual Yang-Mills equations (\ref{3.32}).
\end{theorem}

As a simple partial consequence from Theorem 3.2 one retrieves all of
results \ obtained before in \cite{Gu}, where the Delsarte-Darboux mapping (%
\ref{3.50}) was chosen completely a priori without any proof and motivation
in the form of some affine gauge transformation.

The results similar to the above can be with a minor change applied also to
the affine differential de Rham-Hodge-Skrypnik complex (\ref{3.29}) with the
external differentiation (\ref{3.30}), where
\begin{eqnarray}
\mathrm{L}_{i}(\lambda ) &:&=\mathbf{1}\frac{\partial }{\partial p_{i}}%
-(\sum_{k=0}^{n_{i}(L)}a_{ik}\lambda ^{k+1})\frac{\partial }{\partial x_{i}}%
+\sum_{k=0}^{n_{i}(L)}A_{ik}\lambda ^{k},\text{ \ }  \notag \\
\mathrm{\tilde{B}}(\lambda ) &:&=\partial /\partial t-\sum_{s=0}^{n(B)+q}%
\bar{B}_{s}\lambda ^{n(B)-s},  \label{3.55}
\end{eqnarray}%
or
\begin{eqnarray}
\mathrm{L}_{i}(\lambda ) &:&=\mathbf{1}\frac{\partial }{\partial p_{i}}%
-(\sum_{k=0}^{n_{i}(L)}a_{ik}^{(j)}\lambda ^{k+1})\frac{\partial }{\partial
x_{j}}+\sum_{k=0}^{n_{i}(L)}A_{ik}\lambda ^{k},\text{ \ }  \notag \\
\mathrm{\tilde{B}}(\lambda ) &:&=\partial /\partial t-\sum_{s=0}^{n(B)+q}%
\bar{B}_{s}\lambda ^{n(B)-s},  \label{3.56}
\end{eqnarray}%
for $i=\overline{1,m},$ \bigskip $\lambda \in \mathbb{C}.$ The case (\ref%
{3.55}) was analyzed recently in \cite{We} by means of the same affine gauge
transformation that there was used before in \cite{Gu}. To the regret, the
obtained there results are too complicated and unwieldy, thereby one needs
to use more mathematically motivated, clear and less cumbersome techniques
for finding Delsarte-Darboux transformations and related with them
soliton-like exact solutions.

\section{Acknowledgements}

The authors thank the Organizing Committee of the 36-th Symposium on
Mathematical Physics held June 9-12, 2004 in Torun University, Poland, for \
invitation to deliver the results of the article for Symposium audience. One
of authors (A.P.) cordially thanks Prof. I.V. Skrypnik (IM, Kyiv and IAM,
Donetsk) for fruitful discussions of some aspects of the De Rham
-Hodge-Skrypnik theory and its applications presented in the article during
his Kyiv-city seminar "Nonlinear Analysis" (April 14, 2004). The authors are
appreciated very much to Profs. L.P. Nizhnik (IM of NAS, Kyiv), Holod P.I.
(UKMA, Kyiv), T. Winiarska (IM, Politechnical University, Krakow), A.
Pelczar and J. Ombach (Jagiellonian University, Krakow), J. Janas (Institute
of Mathematics of PAN, Krakow), Z. Peradzynski (Warsaw University, Warsaw)
and D.L. Blackmore (NJ Institute of Technology, Newark, NJ, USA) for
valuable comments on diverse problems related with results presented in the
article. The last but not least thanks is addressed to our friends Profs
V.V. Gafiychuk (IAPMM, Lviv) and Ya. V. Mykytiuk (I.Ya. Franko National
University, Lviv) for the permanent support and help in editing the article.

\end{document}